\documentclass[10pt,a4paper]{article}
\pdfoutput=1
\usepackage{curves,multicol}
\usepackage[export]{adjustbox}
\usepackage{subcaption}
\usepackage[makeroom]{cancel}
\usepackage{scalerel}
\usepackage{natbib}
\usepackage{graphicx}
\usepackage{hyperref}
\usepackage{authblk}
\usepackage{vmargin}
\usepackage{amssymb}
\usepackage{amsfonts}
\usepackage{multirow}
\usepackage{framed}
\usepackage{latexsym}
\usepackage{array}
\usepackage{tensor}
\usepackage{url}
\usepackage{xcolor}
\usepackage{epstopdf, epsfig}
\usepackage{booktabs}
\usepackage{commath}
\usepackage{rotating}
\usepackage{tabularx}
\usepackage{bibentry}
\usepackage{doi}
\usepackage{lineno}
\usepackage{lipsum}
\usepackage{titlesec}
\usepackage{comment}
\newcolumntype{P}[1]{>{\raggedright\arraybackslash}p{#1}}

\providecommand{\keywords}[1]{\textbf{\textit{Keywords ---}} #1}

\newcolumntype{A}{ >{$} r <{$} @{} >{${}} l <{$} } 
\newcolumntype{L}[1]{>{\raggedright\let\newline\\\arraybackslash\hspace{0pt}}m{#1}}
\newcolumntype{C}[1]{>{\tiny\centering\let\newline\\\arraybackslash\hspace{0pt}}m{#1}}
\newcolumntype{R}[1]{>{\raggedleft\let\newline\\\arraybackslash\hspace{0pt}}m{#1}}
\newcolumntype{N}{@{}m{0pt}@{}}

\makeatletter
\newcommand{\smallsym}[2]{#1{\mathpalette\make@small@sym{#2}}}
\newcommand{\make@small@sym}[2]{%
  \vcenter{\hbox{$\m@th\downgrade@style#1#2$}}%
}
\newcommand{\downgrade@style}[1]{%
  \ifx#1\displaystyle\scriptstyle\else
    \ifx#1\textstyle\scriptstyle\else
      \scriptscriptstyle
  \fi\fi
}
\makeatother

\hypersetup{
	unicode=true,
	pdftitle={A descriptive title},
	pdfnewwindow=true,
	colorlinks=true, 	
	pdfborder={0 0 0},
	linkcolor=blue,
	linktoc=all, 		
	citecolor=blue,
	urlcolor=blue,
	breaklinks=false,
}
\setmarginsrb { 0.8in}  
              {0.25in}  
              { 0.8in}  
              { 0.5in}  
              {  20pt}  
              {0.25in}  
              {   9pt}  
              { 0.3in}  

\graphicspath{{img/}}

\usepackage{algpseudocode}
\usepackage{algorithm,setspace}

\newcommand{\algcodesplit}{0.5}

\newcommand{\algalign}[2]{%
  \makebox[\algcodesplit\linewidth][l]{#1}%
  \parbox[t]{\linewidth - \algcodesplit\linewidth}{$\triangleright$ #2}%
}

\title{\textbf{Acceleration of an algebraic multigrid pressure solver using graph neural
networks}}
\author[1,\footnote{\href{mailto:E.ChillonLizana@tudelft.nl}{\texttt{E.ChillonLizana@tudelft.nl}}\newline
Reproducibility repository: \url{https://github.com/ericchillon/AMG-GNN-Jacobi}}]{Eric Chill\'on}
\author[2]{Artur K. Lidtke}
\author[3]{Nguyen Anh Khoa Doan}
\author[1]{Bernat Font}

\affil[1]{Faculty of Mechanical Engineering, Delft University of Technology, The Netherlands}
\affil[2]{Maritime Research Institute Netherlands, The Netherlands}
\affil[3]{Department of Aeronautics, Imperial College London, United Kingdom}
\titlelabel{\thetitle.\,\,}
\date{\normalsize\today}

\begin{document}

{\let\newpage\relax\maketitle}
\setlength{\parindent}{0pt}
\setlength{\parskip}{8pt}

\date{\vspace{-20pt}}
\begin{abstract}
Solving the pressure-Poisson equation remains the primary computational bottleneck in incompressible unstructured flow solvers primarily due to the inherent sensitivity of traditional linear solvers to mesh irregularities. This work introduces a data-driven algebraic multigrid (AMG) smoother that uses a modified graph convolutional isomorphism network (GCIN). The graph neural network predicts optimal polynomial coefficients to construct a sparse pseudo-inverse operator across diverse grid topologies. The coefficients are optimized to reduce the residual after each V-cycle iteration. By directly capturing the algebraic structure of the system from the sparse coefficient matrix, the proposed method maintains the solver's linearity while adapting to local anisotropies in unstructured grids. Our framework demonstrates significant performance gains by reducing the number of V-cycles required for a given tolerance and delivering wall-clock speedups from 4\% to 37\% across diverse benchmarks. Notably, the model exhibits robust generalization by maintaining efficiency on meshes up to 128 times larger than those seen in training, and by accelerating the solver's convergence on unseen industry-relevant problems such as the AirfRANS dataset.
\end{abstract}



\section{Introduction}

Solving the pressure-Poisson equation is a major bottleneck in incompressible flow solvers. At each time step, the discretized Poisson system yields a large, sparse linear system of whose solution dominates computational cost. While momentum equations involve local stencils, the elliptic nature of the pressure-Poisson system requires information to propagate across the entire domain, leading to high communication-to-computation ratios that degrade scaling on massive parallel clusters \citep{Bergman2008,jha2020measuring}. Methods such as asynchronous algorithm communication by \cite{Asynchronous}, tensor-based formulations by \cite{TensorBased}, or communication-reduced conjugate gradients by \cite{CommmRedCG} aim to reduce communication overhead. Nevertheless, the overall speedup is still fundamentally constrained by the efficiency of the underlying linear solver.

Among available solvers, algebraic multigrid (AMG) has become a widely adopted method due to its general applicability and near-linear computational complexity \citep{trottenberg2001multigrid,Baker2011}. However, AMG's performance on complex, unstructured meshes is highly sensitive to the choice of its internal components, such as coarsening strategies, interpolation operators, and smoothers. As a result, achieving robust and efficient performance across different geometries and discretizations remains challenging.

Consequently, data-driven techniques have progressively been explored to accelerate computational fluid dynamics simulations. Early efforts primarily focused on data assimilation and turbulence closures to enable faster, lower-resolution simulations \citep{asch2016data,Wu2018,Maulik2019,Beck2019,Font2021}. More recent work explores broader applications, including machine-learning-guided adaptive meshing \citep{Jha2025} and neural-network surrogates that attempt to replace numerical solvers entirely \citep{NN_muligrid_solver,Ronneberger2015,Seo2024,Le2021}. While promising, these approaches often lack the robustness and convergence guarantees required for high-fidelity simulations. 

Therefore, recent studies have begun investigating machine-learning approaches that accelerate existing solvers rather than replacing them. In the context of the pressure projection step, several works propose learning-enhanced preconditioners to reduce solve times \citep{Li2023,PEARL2025}. However, these approaches are typically restricted to Cartesian grids or specific geometries. Other studies focus on data-driven initial guesses, where an optimized initial guess significantly reduces the residual before the first iteration begins. For instance, \cite{Ozbay2021} uses a convolutional neural network (CNN) to predict cheap solutions of the Poisson equation that can be used as initial guesses. Further, \cite{Vigot2026} uses recurrent GNNs in Poisson equations to generate an initial guess in arbitrary simulation domains. Regarding hybrid approaches, \cite{AjuriaIllarramendiBauerheimCuenot2022} couples a CNN with a traditional iterative solver to ensure a user-defined accuracy level. All in all, while promising, this leaves an important gap in using machine learning to improve core AMG components for general unstructured systems encountered in practical flow simulations.

Recent work on data-driven multigrid methods can broadly be categorized into two directions: (i) the optimization of the grid hierarchy, and (ii) the enhancement of the relaxation process. The first research direction focuses on improving coarsening and interpolation operators. For instance, restriction operators have been optimized by learning suitable strength-of-connection parameters. Selecting appropriate strength parameters can significantly improve convergence rates, thereby reducing the number of AMG iterations. This has been done using multilayer perceptron (MLP) neural networks \citep{Caldana2019}, CNNs \citep{Caldana2024}, and graph neural networks (GNNs) \citep{HaifengZou2024}, with the latter offering greater flexibility for unstructured meshes. CNNs are unsuitable for unstructured meshes because they rely on fixed kernels and assume similar spatial relationships repeat across the grid, which is not true for varying element shapes (e.g., tri-quad meshes). Other approaches construct coarsening operators using reinforcement learning agents, which can improve memory efficiency, although sometimes at the expense of slower convergence rates \citep{Taghibakhshi2021}. Similarly, several studies have explored learning prolongation operators using GNNs such as \cite{Greenfeld2019,Luz2020,Wang2023}, demonstrating promising generalization properties on unseen unstructured problems. Additional approaches improve the coarse-grid correction by enhancing aggregation-based AMG with GNNs \citep{Nytko2022} or generating more efficient Galerkin coarse-grid operators \citep{Huang2024} in unstructured and structured meshes, respectively. Overall, improvements in the grid hierarchy have shown consistent benefits, with strength-parameter tuning often among the most effective strategies.

The second research direction focuses on accelerating the smoothing step of multigrid methods. Although less extensively studied, improving the smoother can have a substantial impact on overall solver performance. For example, \cite{kuznichov2022learningrelaxationmultigrid} uses a neural network to predict optimal relaxation parameters for relaxed Jacobi and multicolor Gauss–Seidel smoothers, achieving faster convergence than conventional approaches. \cite{Huang2023} replaces the smoother entirely with a convolutional neural network that learns to reduce residual errors directly from structured matrix representations. Similarly, \cite{Weymouth2022} proposes a Jacobi-based smoother derived from a pseudo-inverse operator with the same sparsity pattern as the Poisson matrix, demonstrating significant improvements in geometric multigrid solvers.

Despite these advances, several limitations remain in current data-driven multigrid research. Most existing methods are developed and validated primarily on structured grids, often in two-dimensional settings. Furthermore, many studies focus on reducing iteration counts rather than demonstrating improvements in overall wall-clock solve time, which ultimately determines practical efficiency. Finally, relatively few works evaluate their methods on complex, application-relevant problems, leaving open questions regarding robustness, scalability, and practical applicability in industry-scale simulations.

Building on this foundation, this work introduces a data-driven smoother for AMG, extending the work of \cite{Weymouth2022} for unstructured grids.  This framework predicts optimal polynomial coefficients for constructing a sparse pseudo-inverse smoother using a GNN. By employing a modified graph convolutional isomorphism network (GCIN) architecture \citep{HaifengZou2024}, the model directly captures the algebraic structure of the pressure-Poisson system from the sparse coefficient matrix while preserving the solver's underlying linearity. This framework achieves significant performance gains, reducing V-cycle iterations by up to 36\% and delivering wall-clock speedups of 17\% to 25\% across diverse benchmark problems. Notably, the GNN learns generalized algebraic patterns that extend beyond specific flow configurations, maintaining V-cycle reduction generalization on unseen industry-relevant problems, such as the AirfRANS dataset \citep{bonnet2023airfranshighfidelitycomputational}, and on mesh resolutions up to 128 times larger than those encountered during training.

The remainder of this paper is organized as follows. Section \ref{sec:2} establishes the mathematical foundation of the discretized pressure-Poisson equation and explores the specific impact of smoother selection and parameter sensitivity on algebraic multigrid (AMG) convergence. Section \ref{sec:3} introduces our data-driven enhanced smoother framework, detailing the construction of the sparse pseudo-inverse operator, the modified graph convolutional isomorphism network (GCIN) architecture, and the end-to-end residual reduction loss function used for training. In Section \ref{sec:4}, we evaluate the performance, robustness, and generalization of the proposed framework across a series of numerical experiments, including synthetic problems on structured grids, canonical laminar flows on unstructured meshes, and high-fidelity aerodynamic simulations using the industry-relevant AirfRANS dataset. Finally, Section \ref{sec:5} provides concluding remarks, summarizes our core findings, and outlines current limitations and directions for future research.

\section{Smoother sensitivity}\label{sec:2}

\subsection{Sparse linear system}

The discretized pressure-Poisson equation, which yields a large and extremely sparse linear system during each time step of incompressible flow solvers, is defined as
\begin{equation}\label{eq:poissonequation}
    \mathbf{A}\mathbf{u} = \mathbf{f},
\end{equation}
where $\mathbf{u}$ is the unknown pressure vector, $\mathbf{f}$ is the source term vector, $\mathbf{A} \in \mathbb{R}^{n \times n}$ is the discrete matrix of the Poisson operator, and $n$ is the number of unknowns. In the context of conservative Poisson operators, $\mathbf{A}$ is symmetric, has zero-sum rows and columns, is negative semi-definite, and has a single zero eigenvalue. The presence of a single zero eigenvalue corresponds to the null space of the constant vector. Physically, this reflects that the pressure in an incompressible flow is only defined up to an additive constant.

Due to the scale of these systems, iterative methods are preferred over direct methods. These processes iteratively refine the solution $\mathbf{u}^{k+1}$ by dampening the algebraic error $\boldsymbol{\epsilon}^k = \mathbf{u} - \mathbf{u}^k$. In practice, the solution is updated by an error approximation $\boldsymbol{\epsilon}^k$:
\begin{equation}
\mathbf{u}^{k+1} = \mathbf{u}^k + \boldsymbol{\epsilon}^k.
\end{equation}
Given the linearity of the problem, the update $\boldsymbol{\epsilon}$ is determined from
\begin{equation}\label{eq:updateequation}
    \mathbf{A}\boldsymbol{\epsilon}^k = \mathbf{r}^k \equiv \mathbf{f} - \mathbf{A}\mathbf{u}^k.
\end{equation}
In the context of stationary iterative methods, the error approximation is often computed as $\boldsymbol{\epsilon}^k = \mathbf{Q}^{-1}\mathbf{r}^k$, where $\mathbf{Q}$ is an easily invertible operator that approximates $\mathbf{A}$. The choice of $\mathbf{Q}$ is the fundamental factor in determining the convergence rate of the solver.

\subsection{AMG algorithm}

The algebraic multigrid algorithm relies on a setup and a solve phase.  During the setup phase of a two-level AMG, the algorithm constructs a coarse level with associated restriction and prolongation matrices $\mathbf{R}$ and $\mathbf{P}$ based on the coefficient matrix $\mathbf{A}$. The coarse-level grid depends on the specific coarsening algorithm and parameters, such as the strength threshold $\theta$. Further, the coarse level matrix is then determined as $\mathbf{A}_c = \mathbf{R}\mathbf{A}\mathbf{P}$. This construction, known as the Galerkin triple product, ensures that the coarse operator is the best representation of the fine operator within the range of the interpolation matrix $\mathbf{P}$. In larger hierarchies, the algorithm is applied recursively until the coarsest level can be easily solved by a direct solver.

In the solve phase, a standard multigrid cycle is performed based on the matrices generated in the setup phase. A two-level AMG algorithm involves (i) pre-smoothing, (ii) restricting residuals to the coarse level, (iii) solving the residual equations in the coarse level, (iv) interpolating the error approximation back to the fine level, and (v) post-smoothing. For hierarchies with more than two levels, the direct solve is reserved for the coarsest level. The sequence in which the algorithm traverses the grid hierarchy determines the cycle type, with the V-cycle being the most common standard.

AMG performance on unstructured meshes is highly sensitive to the choice of these internal components. Much of the research in AMG has focused on the setup phase, specifically in optimizing the interpolation operator $\mathbf{P}$ \citep{Greenfeld2019,Luz2020,Wang2023}. However, as the system grows and is deployed on parallel architectures, the efficiency of the smoother becomes a critical bottleneck for total wall-clock time.

\subsection{Smoother impact in convergence}

The efficacy of the AMG method relies on the complementary relationship between the grid hierarchy and the relaxation process. Smoothers are specifically designed to eliminate high-frequency (oscillatory) errors that cannot be well-represented on coarser grids. The efficiency of this process is quantified by the smoothing factor, which represents the reduction in oscillatory error per iteration. Once these local errors are dampened, the remaining low-frequency (smooth) errors are effectively handled by the multigrid hierarchy.

While basic smoothers like relaxed Jacobi or multicolor Gauss-Seidel can reduce high-frequency error, they often become a limiting factor in overall wall-clock solve time. This introduces a trade-off between the mathematical strength of the smoother and its computational efficiency. For example, while Gauss-Seidel typically provides superior error reduction per step, its sequential dependencies can hinder performance on parallel computing backends. In contrast, the relaxed Jacobi smoother, despite requiring more V-cycles, offers a lower per-iteration cost and better scalability, potentially resulting in faster overall solve times.

The speed of convergence is highly sensitive to the relaxation parameter $\omega$, which determines the weight of the iterative correction. For instance, in relaxed Jacobi
\begin{equation}
\mathbf{u}^{k+1} = \mathbf{u}^{k} + \omega \mathbf{Q}^{-1}\mathbf{r}^k = \mathbf{u}^{k} + \omega \boldsymbol{\epsilon}^k.
\end{equation}
On complex and unstructured meshes, the performance of standard smoothers often degrades. A fixed, global $\omega$ fails to account for local algebraic variations and anisotropies inherent in unstructured discretizations, making it difficult to uniformly dampen high-frequency errors across the entire domain. This highlights the need for smoother formulations that can adapt to the local structure of the coefficient matrix at each grid level.

\section{GNN-based smoother}\label{sec:3}

\subsection{Pseudo-inverse smoother}\label{sec:3.1}

The enhancement of the algebraic multigrid solver centers on developing a data-driven Jacobi smoother. Rather than utilizing a standard relaxation parameter or the simple inverse of the diagonal as in classical implementations, this work constructs a sparse pseudo-inverse smoother matrix, $\mathbf{Q}^{-1}$ (or $\tilde{\mathbf{A}}^{-1}$), with the same sparsity pattern as the original coefficient matrix $\mathbf{A}$.

A key advantage of this formulation is that the linearity of the underlying operator $\mathbf{A}$ is preserved. Non-linearities are confined strictly to the construction of the smoother coefficients $\mathbf{Q}^{-1}$ via the GNN. Therefore, the convergence properties of the solver are maintained. Other methods, such as enhancing prolongation and restriction operators, also maintain the linearity. Additionally, when $\mathbf{A}$ is constant during the simulation, the pseudo-inverse can be computed and stored ahead of time. This incurs a marginal one-time overhead that is rapidly recovered by the reduction in V-cycle counts.

The pseudo-inverse is constructed using polynomial functions whose coefficients are predicted by a GNN. Specifically, the smoother is defined into two parts: the diagonal and the off-diagonal. To build the smoother matrix, we follow \cite{Weymouth2022}, which is based on structured grids and geometric multigrid, but is generalizable to AMG.
\begin{equation}
q_{ii}^{-1} = \frac{f_d(a_{ii}/s \ | \ \boldsymbol{\theta})}{a_{ii}}, \quad q_{ij}^{-1} = \frac{f_o(a_{ij}/s \ | \ \boldsymbol{\theta})}{a_{ii} + a_{jj}} \quad (i \neq j).
\end{equation}
Here $s = \max|\mathbf{A}-\mathbf{D}|$ is a scaling factor for the inputs, and $f_d$, $f_o$ are second-degree polynomial functions. $\mathbf{D}$ is the diagonal of matrix $\mathbf{A}$. Note that $f_o(0) = 0$ is required to maintain the sparsity. The second-degree choice is critical. We have found that higher-order polynomials severely overfit to specific mesh topologies, leading to instabilities when generalized to unseen unstructured grids.

\subsection{GNN architecture}

\begin{figure}[h!]
    \centering
    \includegraphics[width=0.95\linewidth]{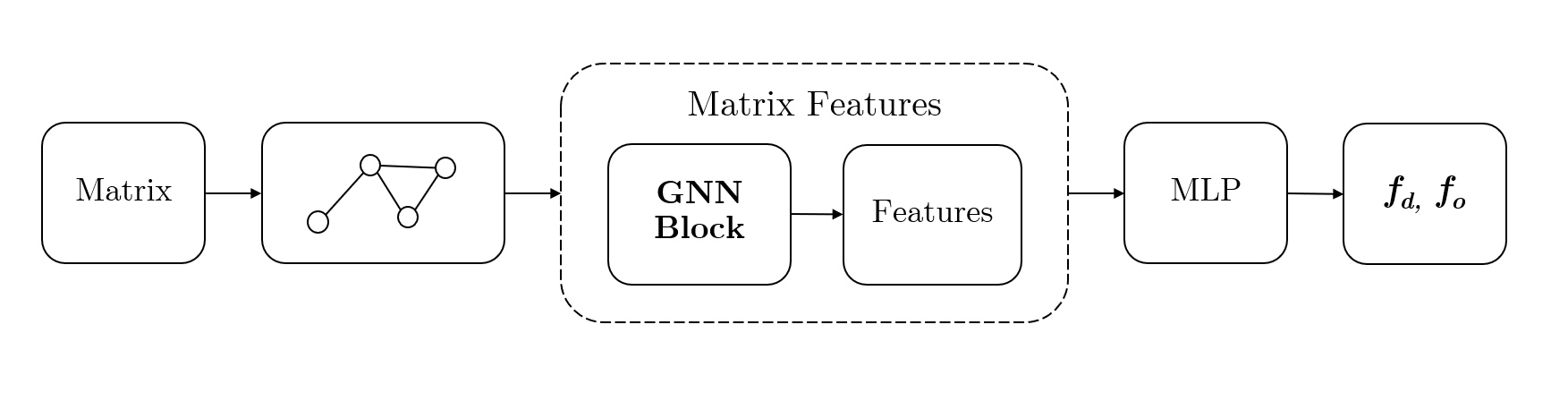}
    \caption{GNN-based smoother method.}
    \label{fig:gnnscheme}
\end{figure}

To predict the optimal coefficients for the pseudo-inverse, we use a modified GCIN for feature extraction and an MLP network to predict the coefficients, as shown in Figure \ref{fig:gnnscheme}. These are determined at each hierarchy level, yielding a tailored, data-driven smoother for each level. This architecture, inspired by AutoAMG \cite{HaifengZou2024}, combines a graph convolutional network (GCN) and a graph isomorphism network (GIN). GCNs operate by taking a weighted average of their neighbors' features, making them highly efficient at extracting structural patterns from irregular domains such as unstructured grids. On the other hand, GINs upgrade this by using a sum-aggregation mechanism, providing the mathematical expressiveness needed to distinguish complex, distinct graph topologies without washing out structural nuances. Combined, they enable the model to accurately capture localized mesh variations and anisotropies across diverse grid layouts, thanks to the efficiency of GCNs and the expressiveness of GINs. The GCIN is suited for unstructured meshes because it operates directly on the graph representation of the sparse matrix $\mathbf{A}$, where nodes represent grid cells and edges represent their algebraic connections. 

The coefficient matrix is represented as a graph $G$, defined as a tuple $(V, E)$, consisting of a set of nodes $V$ and a set of edges $E$. The cardinalities $|V|$ and $|E|$ represent the total number of nodes and edges, respectively. Within this framework, $v_i \in V$ identifies the $i$-th node, while a directed edge from $v_i$ to $v_j$ is denoted by $e_{ij} = (v_i, v_j) \in E$. The set of neighbors for any node $v_i$ is represented by $N(v_i)$. To account for node and edge attributes, $\mathbf{X}_v \in \mathbb{R}^{|V| \times d}$ and $\mathbf{X}_e \in \mathbb{R}^{|E| \times c}$ are introduced as the respective feature matrices. Here, $\mathbf{x}_{v_i} \in \mathbb{R}^d$ corresponds to the feature vector of node $v_i$, and $\mathbf{x}_{e_{ij}} \in \mathbb{R}^c$ corresponds to the feature vector of edge $e_{ij}$.

The matrix $\mathbf{A}$ is mapped to a graph as follows. Each node has a two-dimensional feature vector. The stored values are the diagonal coefficient of the matrix and the node's number of connections (or degree). Adding the node degree provides the GNN with the number of mesh cells connected to a given cell. That is, whether the cell is 2D or 3D, and the type of element.
\begin{equation}
    {\mathbf{x}_v}_i^{(0)} = \begin{pmatrix}
a_{ii} \\
\text{deg}(v_i) 
\end{pmatrix}
\end{equation}
The edges are assigned a scalar attribute, equal to the off-diagonal coefficient $w_{ij} = a_{ij}$. They represent the spatial connection between two cell nodes $i$ and $j$. Further, large weights imply stronger connections between the cells. Given that we are interested in the node outputs, this is a node-based network. Therefore, the input feature dimension is equal to 2 in the first GCIN layer.

A GCIN is constructed on the GCN. To add the expressiveness inherent to a GIN, normalization within the GCN is no longer performed, avoiding feature cancellation from averaging and increasing the range of values the graph network block sees. Hence, the forward pass of the GCIN for a layer $l$ is defined as
\begin{equation}\label{eq:GCN}
    \mathbf{x}_{v_i}^{(l)} = \text{MLP}^{(l)} \left( \sum_{v_j \in N(v_i)\cup v_i} w_{ij}\mathbf{x}_{v_j}^{(l-1)}\right),
\end{equation}
where $w_{ij}$ is the weight of the edge $e_{ij}$. Note that the edge feature vectors are simplified to scalars in this work. Further, its matrix form is
\begin{equation}
    \mathbf{x}_v^{(l)} = \hat{\mathbf{A}}\mathbf{X}_v^{(l-1)}\boldsymbol{\Theta}^{(l)}.
\end{equation}
Here $\boldsymbol{\Theta}^{(l)}$ is the weight matrix for the $\text{MLP}^{(l)}$ that is optimized during training, and $\hat{\mathbf{A}}$ is the adjacency matrix of the graph with added self-loops, so each node is connected to itself.

The architecture consists of sequential GCIN layers. By stacking multiple GCIN layers, information from farther neighbors is captured, as each GCIN pass convolves the first-hop neighbors. The outputs from the layers are globally pooled (mean, max) and concatenated before the predictive MLP, from which optimal polynomial coefficients are inferred. Therefore, the readout function for the GCIN is
\begin{equation}
    \mathbf{x}_{g}^{(l)} = \text{CONCAT} \left[ \frac{1}{|V|} \sum_{i=1}^{|V|} \mathbf{x}_{v_i}^{(l)}, \max_{i=1, \dots, |V|} \left( \mathbf{x}_{v_i}^{(l)} \right) \right],
\end{equation}
\begin{equation}
    \mathbf{x}_{G} = \text{CONCAT}\left(\mathbf{x}_{g}^{(l)} \ \bigg | \ l = 0,1, \dots, L\right).
\end{equation}

The GNN model comprises four sequential layers ($L=4$). The first GCIN layer input dimension is 2. The following layers present hidden and output dimensions of 64. Hence, the extracted matrix features are of size $128$ due to the double global pooling. The extracted features are concatenated into an array of size $128L$, and introduced into the MLP. The latter approximates the mapping between matrix features and the optimal polynomial coefficients. The MLP consists of two layers. The input dimension is $128L$, the hidden dimension is $128$, and the output dimension is 5. The latter is due to the independent off-diagonal coefficient being strictly zero ($f_o(0) = 0$). Hence, there are 3 coefficients for the diagonal polynomial and two for the off-diagonal. This proved to be the best network size based on a preliminary hyperparameter study (Appendix \ref{app:hyperparam_opt}). Detailed information about architectural hyperparameters is presented in Appendix \ref{app:hyperparam_arch}.

\subsection{Optimizing residual reduction}

The GNN is trained end-to-end across the entire AMG V-cycle using PyTorch's automatic differentiation. Unlike methods that focus on the spectral radius of the error propagation matrix, which can be computationally prohibitive to evaluate for large, diverse datasets, this work employs a residual-reduction loss function. By avoiding the calculation of spectral attributes, such as the condition number or the eigenvalue distribution, the training process remains scalable to 3D systems.

Optimizing for residual reduction has several practical advantages. Primarily, it directly influences the total wall-clock time. Decreasing the number of V-cycles reduces the total computational effort, provided the smoother construction cost is managed. Second, the residual is always available, unlike other metrics such as the error with respect to the unknown exact pressure solution. Hence, no true solution is necessary for training. Finally, since every step of the AMG V-cycle (including smoothing, restriction, and prolongation) consists of differentiable linear operations, the GNN can be trained end-to-end. This allows the network to learn how its smoother coefficients affect the global error reduction across the entire grid hierarchy, rather than just optimizing for a single local relaxation step.

The training objective is to minimize the $L_2$ norm of the residual after one or more V-cycles ($N_{vc}$). The primary loss function is defined as
\begin{equation}
    L = \sum_{n=1} ^{N_{vc}}\log_{10}\left( \dfrac{||\mathbf{r}_{n}||_2}{||\mathbf{r}_{n-1}||_2}   \right).
\end{equation}

This logarithmic formulation makes it easy to monitor performance, as a negative loss directly indicates that the learned smoother is effectively reducing the residual. To ensure the model generalizes beyond the first iteration, a multi-V-cycle (MVC) loss can be used to average reductions across multiple iterations, which can be beneficial for very specific problems. 

The optimal parameters of the accelerated AMG projection method are then given by the minimization
\begin{equation}
    \hat{\boldsymbol{\theta}} = \min_{\boldsymbol{\theta}} \ L(\boldsymbol{\theta} \ | \ \mathcal{T} )
\end{equation}
where $\mathcal{T} = \{\mathbf{A},\mathbf{f},\mathbf{u}_0\}$ is the training dataset and $\boldsymbol{\theta}$ are the learnable parameters from the GNN and the predictive MLP.

During training, the model's performance is monitored against a baseline relaxed Jacobi smoother. This relative validation metric, defined as
\begin{equation}
    L = \log_{10}\left( \dfrac{||\mathbf{r}_{n+1}||_2}{||\mathbf{r}^{\omega -J}_{n+1}||_2}   \right)
\end{equation}
serves as an early stopping criterion, ensuring that the final model provides a measurable improvement over conventional stationary methods.

\subsection{Training and testing procedures}

The algebraic multigrid level hierarchy varies significantly depending on the coefficient matrix and, hence, the mesh. This AMG aspect becomes a limitation when handling mini-batches consisting of different problems. As a result, while parallelizable, the AMG solver does not admit batched V-cycles unless the mini-batches are constructed in such a way that the AMG hierarchies are identical. That is, the same number of levels and coarse matrix sizes, which can involve considerable data manipulation. Further, using sparse matrices makes the issue harder to handle. Therefore, we opted for a sequential training V-cycle implementation. 

The sequential approach applies to both model training and evaluation. Regarding training, following this approach slows the process, but it avoids hindering stochasticity during learning. Regarding testing, solving the Poisson systems individually is a more natural approach. This allows to smoothly evaluate setup and solve times without performance variations due to batched solving.

To optimize the training, given the sequential V-cycle limitation, we use an initial batched GNN inference. Then, the predicted batch of polynomial coefficients is used to individually construct the hierarchy of pseudo-inverse smoother matrices for the V-cycle. The process is presented in Algorithm \ref{alg:gnn_amg_training}. Further details on training and testing can be found in Appendix \ref{app:train_procedure} and \ref{app:test_procedure}, respectively.

\begin{algorithm}[h!]
\setstretch{1.2}
\caption{Training and inference of the GNN-based AMG Smoother}
\label{alg:gnn_amg_training}
\begin{algorithmic}[1]
    \Function{GNN-AMG-Solver}{$\mathcal{T}= \{\mathbf{A}, \mathbf{f}, \mathbf{u}_0\}$, $\boldsymbol{\theta}$}
        \For{each mini-batch $\mathcal{B} \subset \mathcal{T}$}
            \State \algalign{\text{Initialize batch graph container } $\mathcal{G}_{batch} \gets \emptyset$}{1. Prepare batch structures}
            \For{each problem $m \in \{1 \ldots |\mathcal{B}|\}$}
                \State \algalign{$\{\mathbf{A}^{(\ell)}, \mathbf{R}^{(\ell)}, \mathbf{P}^{(\ell)}\}_{m} \gets \text{Setup-AMG-Hierarchy}\big(\mathbf{A}_m\big)$}{2. Construct hierarchies per problem}
                \For{each level $\ell \in \{0 \ldots \mathcal{L}_m-1\}$}
                    \State \textbf{let} $G_m^{(\ell)} = (V, E)$ represent $\mathbf{A}_m^{(\ell)}$ 
                    \State \algalign{$\mathbf{x}_{v_i}^{(0)} \gets \big[{(a_m)}_{ii}, \text{deg}(v_i)\big]^T, \forall v_i \in V$}{3. Initialize node features}
                    \State \algalign{$\mathbf{x}_{e_{ij}} \gets {(a_m)}_{ij}, \forall (i,j) \in E$}{4. Initialize edge attributes }
                    \State \algalign{$\mathcal{G}_{batch} \gets \mathcal{G}_{batch} \cup G_m^{(\ell)}$}{5. Collect all levels into batch}
                \EndFor
            \EndFor
            \Statex 
            \State \algalign{$\hat{\mathbf{P}}_{batch} \gets \text{GNN}_{\boldsymbol{\theta}}(\mathcal{G}_{batch})$}{6. Batched GNN inference}
            \Statex 
            \State \algalign{$L_{\mathcal{B}}(\boldsymbol{\theta}) \gets 0$}{7. Initialize batch loss}
            \For{each problem $m \in \{1 \ldots |\mathcal{B}|\}$}
                \For{each level $\ell \in \{0 \ldots \mathcal{L}_m-1\}$}
                    \State \algalign{$(\mathbf{Q}^{-1})_m^{(\ell)} \gets \text{AssembleSmoother}\big(\mathbf{A}_m^{(\ell)}, \hat{\mathbf{p}}_m^{(\ell)}\big)$}{8. Construct level-wise smoothers}
                \EndFor
    
                \For{$n \in \{1 \ldots N_{vc} \}$}
                    \State \algalign{$(\mathbf{u}_m)_n \gets \text{V-Cycle}\Big( \{\mathbf{u}_{n-1},\mathbf{A}, \mathbf{R}, \mathbf{P}, \mathbf{Q}^{-1}\}_m\Big)$}{9. Sequential V-cycles per problem}
                \EndFor
                \State \algalign{$\mathbf{r}_n \gets \mathbf{f}_m - \mathbf{A}_m (\mathbf{u}_m)_n$}{10. Compute problem residual}
                \State \algalign{$L_m \gets \sum \log_{10} \big(\|\mathbf{r}_n\|_2 / \|\mathbf{r}_{n-1}\|_2\big)$}{11. Calculate problem loss}
                \State \algalign{$L_{\mathcal{B}} \gets L_{\mathcal{B}} + L_m/|\mathcal{B}| $}{12. Accumulate mean batch loss}
            \EndFor
            \State \algalign{$\boldsymbol{\theta} \gets \text{Optimizer}\big(\nabla_{\boldsymbol{\theta}} L_{\mathcal{B}}(\boldsymbol{\theta}) \big)$}{13. Update weights}
        \EndFor
        \State \textbf{return} Optimized parameters $\boldsymbol{\theta}$
    \EndFunction
\end{algorithmic}
\end{algorithm}


\section{Numerical experiments}\label{sec:4}

Three sets of cases have been used to establish the effectiveness of the GNN-based AMG Jacobi solver: synthetic and unsteady flow on structured meshes, canonical flows on unstructured meshes, and airfoil flows on structured meshes (AirfRANS). The first set aims to compare the AMG with a robust, data-driven geometric multigrid (GMG) (Section \ref{sec:struct}). The other two sets are used to evaluate the model on more relevant and practical problems (Sections \ref{sec:fulltrain}, \ref{sec:halftrain}, and Section \ref{sec:airfrans}).

The linear equations are solved using the custom AMG with a tailored data-driven smoother for each level. The solver uses 2 pre- and post-smoothing steps ($\nu_1$, $\nu_2 = 2$) at each level. The coarsening algorithm is the classical Ruge-Stüben, with strength parameter $\theta = 0.25$. An upper limit of 500 iterations is used, and the solve tolerance is $\delta = 10^{-4}$ following standard RANS CFD practices.

We evaluate the performance of the data-driven smoother against a relaxed Jacobi smoother with relaxation parameter $\omega = 2/3$. This baseline smoother offers three advantages over sequential smoothers such as Gauss-Seidel or SOR. First, it is inherently parallelizable (similarly to the data-driven smoother), and the solution update is obtained by simple matrix multiplication. Second, although it is not as effective as the other iterative methods as a standalone solver, the performance as a smoother is similar. This is due to the AMG hierarchy handling the low-frequency errors. Finally, the PyTorch implementation for GPU acceleration is more straightforward for the relaxed Jacobi than for the sequential GS and SOR. 

In the following plots, the terms acceleration and convergence are used. The acceleration, or speedup, is the average solve time of the relaxed Jacobi AMG divided by the average solve time of the GNN-based Jacobi smoother, also called data-driven Jacobi AMG. Convergence indicates whether the solver reaches a solution. That is, no residual divergence or stagnation.

\subsection{Structured meshes}\label{sec:struct}
The structured dataset consists of synthetic, unsteady flow problems that have already been used as benchmarks in \cite{Weymouth2022}. This data is generated with WaterLily, an incompressible flow solver written in Julia \citep{WaterLily}. The synthetic cases aim to establish the ability of the accelerated AMG to reduce residuals at domain boundaries (static), within the fluid (dipole), and at body boundaries (sphere), in two and three dimensions. The unsteady incompressible flow cases are variations on standard unsteady flow benchmarks, such as 2D flow past a static circular cylinder, 3D flow past a static donut, and the 3D Taylor--Green vortex (TGV) case. For more information, see \cite{Weymouth2022}.

The benchmark evaluation is built upon six distinct categories of synthetic flow problems. For each category, we generated 250 problem instances, each using an identical coefficient matrix $\mathbf{A}$ and a unique source term vector $\mathbf{f}$ generated via random field sampling. The name and grid dimensions for each problem category are detailed in Table \ref{tab:synth_problems}. 

\begin{table}[htbp]
\centering
\caption{Synthetic problems configurations and sizes.}
\label{tab:synth_problems}
\begin{tabular}{lcccc}
\toprule
\textbf{Name}  & \textbf{Training grid size} & \textbf{Testing grid size} & \textbf{Total problems} \\
\midrule
 2D-static & $32^2$ & $128^2$ & 250 \\
 3D-static & $16^3$ & $64^3$  & 250 \\
 2D-dipole & $32^2$ & $128^2$ & 250 \\
 3D-dipole & $16^3$ & $64^3$  & 250 \\
 2D-sphere & $32^2$ & $128^2$ & 250 \\
 3D-sphere & $16^3$ & $64^3$  & 250 \\
\bottomrule
\end{tabular}
\end{table}

Of the 250 total problem instances per category, 200 are allocated for training and 50 are reserved for testing. We utilize this data to train two distinct types of GNN models:

\begin{enumerate}
    \item \textbf{Specific models:} six separate GNNs are trained, with each model restricted exclusively to the 200 training problems of a single category.
    \item \textbf{Union models:} three broader models are trained on mixed datasets combining different problem types. The 2D-Union and 3D-Union models are each trained on 300 problems, drawing an even split of 100 problems from each corresponding dimensional category. The cross-dimensional Union model is trained on 360 problems, with an even split of 60 training problems per category.
\end{enumerate}

\begin{figure}[h!]
    \centering
    \begin{subfigure}[t]{0.49\linewidth} 
        \centering
        \includegraphics[height = 5cm, trim={0pt 2pt 0pt 0pt}, clip]{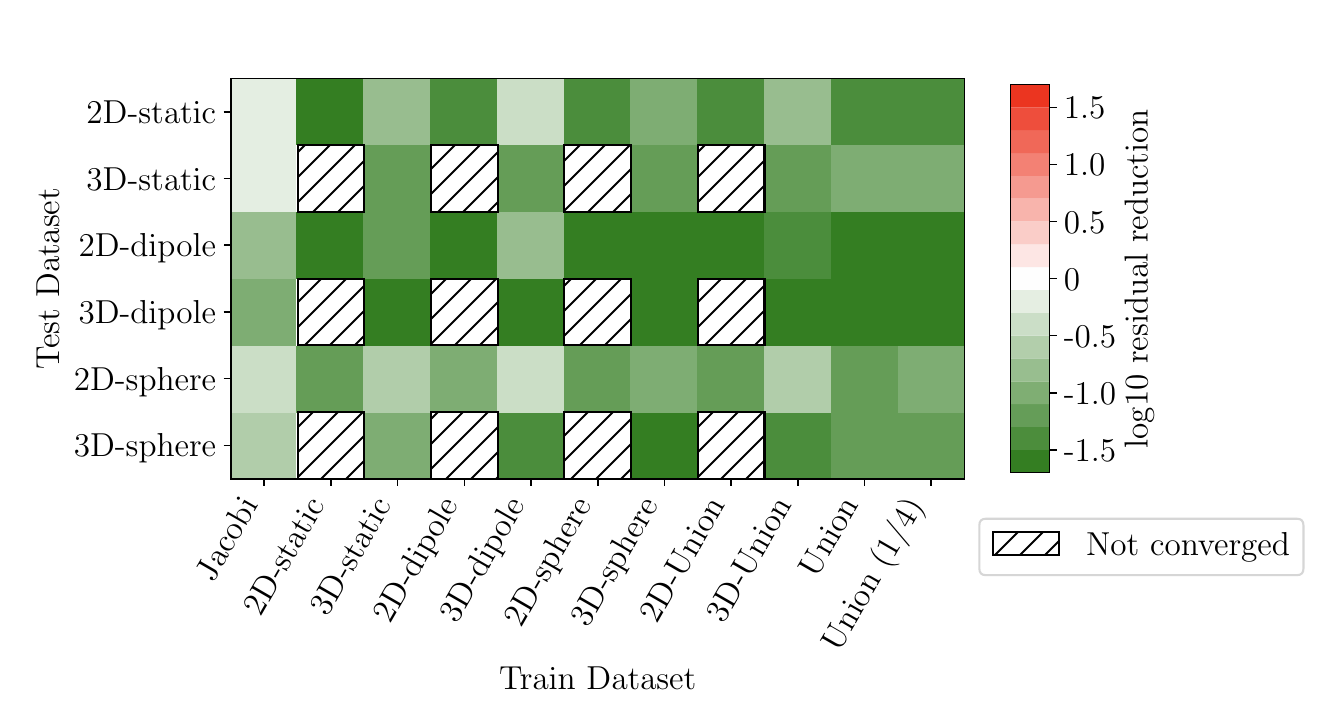}
        \caption{}
        \label{fig:residual.a}
    \end{subfigure}
    \hfill
    \begin{subfigure}[t]{0.49\linewidth} 
        \centering
        \includegraphics[height = 5.1cm, trim={0pt 0pt 0pt 2pt}, clip]{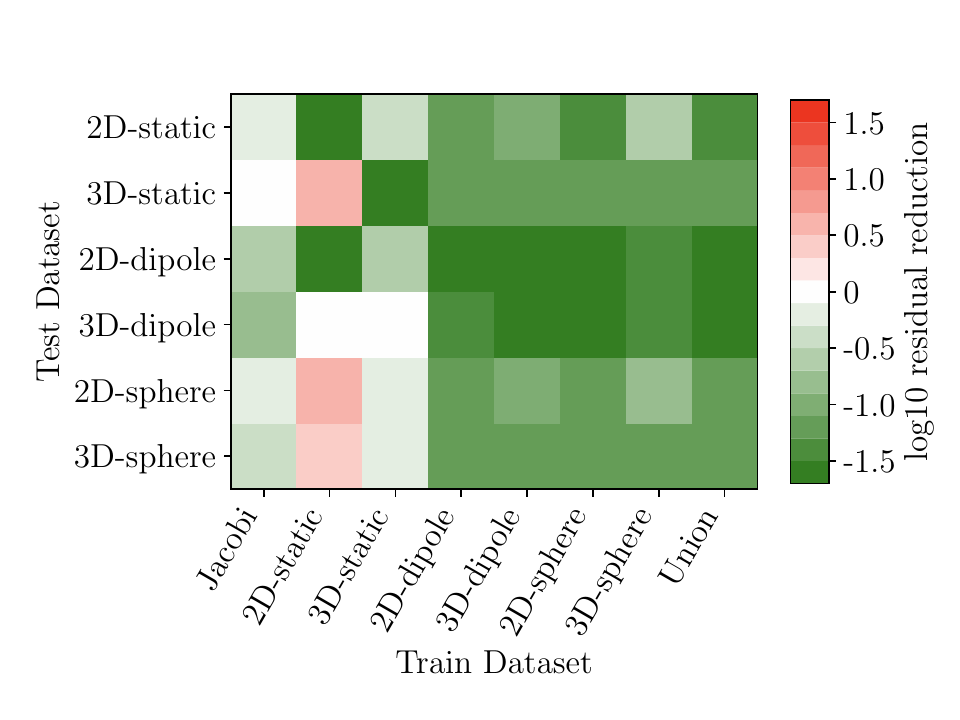}
        \caption{}
        \label{fig:residual.b}
    \end{subfigure}
    \caption{Residual reduction after a V-cycle for (a) GNN-based smoother and (b) data-driven GMG adapted from \cite{Weymouth2022}.}
    \label{fig:residual}
\end{figure}

During the testing phase, each of these trained models is evaluated against the 50 reserved test problems across all six categories. All tests are computed in double precision (FP64) on meshes with a 4x higher resolution per dimension than those used in training (e.g., transitioning from $32^2$ to $128^2$). Figure \ref{fig:residual} illustrates the resulting $\log_{10}$ residual reduction, $\log_{10} \|r\|$, where each data point reflects the mean reduction achieved across the 50 tested problems within that specific problem category.

Looking at the residual reduction performance across different training configurations in Figure \ref{fig:residual.a}, a key limitation of the AMG-GNN framework is its inability to generalize to higher-dimensional structures than those encountered during training. Specifically, models trained exclusively on 2D categories fail to reduce the residual when evaluated on 3D test categories, resulting in solver divergence. This behavior stems from the graph representation of the grid layout. In 3D structured grids, the GNN must convolve across $6+1$ graph nodes (representing the center cell and its six orthogonal neighbors), whereas it is optimized for only $4+1$ nodes in 2D configurations. As a result, the model applies random weights to the extra node connections, producing non-optimal polynomial coefficients. In contrast, an AMG-GNN model trained on a 3D dataset successfully generalizes to unseen 2D categories. This asymmetric generalization demonstrates that the network is truly learning the underlying algebraic patterns of the coefficient matrices rather than mesh- or dimension-specific geometry. 

Beyond matrix structural properties, the data-driven smoother demonstrates a secondary capability to specialize based on flow physics and the right-hand side ($\mathbf{f}$) source terms. This effect is noticeable when the matrix $\mathbf{A}$ is kept constant. The 3D-dipole test category experiences a significantly larger residual reduction than the 3D-static category, highlighting that the model adjusts to specific flow features indirectly encoded within the residual loss function. Because the solver adapts to both layout matrix properties and flow variations, incorporating diverse data during the training phase yields the most robust performance. Training on the combined Union datasets consistently delivers optimal residual reduction across all evaluated test categories. Furthermore, training on smaller problem instances, such as models optimized at a 1/4-th spatial scale relative to the test domain (model Union (1/4) in Figure \ref{fig:residual.a}), retains almost identical error-reduction capability, confirming strong mesh-size generalization.

When comparing these trends directly to the data-driven GMG approach depicted in Figure \ref{fig:residual.b}, both multigrid methodologies demonstrate remarkably similar performance profiles. Despite utilizing distinct grid hierarchies, where the GMG approach is fundamentally bounded by geometric grid restrictions and the AMG-GNN operates via completely automated algebraic coarsening, both models achieve equivalent residual reduction steps when given diverse training datasets. This indicates that the graph neural network's localized coefficient selection matches the convergence efficiency of a tailored geometric multigrid smoother and extends the methodology to irregular, unstructured matrix layouts.

Having established the residual-reduction capabilities on synthetic data, we now assess the performance of the AMG-GNN model on more practical and physically complex unsteady incompressible flows. Based on our previous observations that data diversity consistently provides superior error reduction across all test cases, we again adopt a comprehensive union model approach for training. This choice is crucial for fixed-body unsteady simulations. Because the spatial discretization matrix $\mathbf{A}$ remains constant throughout the simulation, a category-specific model can easily overfit the constant matrix structure and the transient noise of the unsteady flow vectors $\mathbf{f}$, severely hindering its ability to generalize to larger grids. To build this robust training dataset, we compile a union dataset of 900 problems, evenly distributed across 225 instances for each of the following configurations: 2D problems on $16^2$ meshes, 2D problems on $32^2$ meshes, 3D problems on $8^3$ meshes, and 3D problems on $16^3$ meshes. Then, we assess the performance of 3 trained models.

\begin{figure}[h!]
    \centering
    \begin{subfigure}[t]{0.46\linewidth} 
        \centering
        \includegraphics[width = \linewidth]{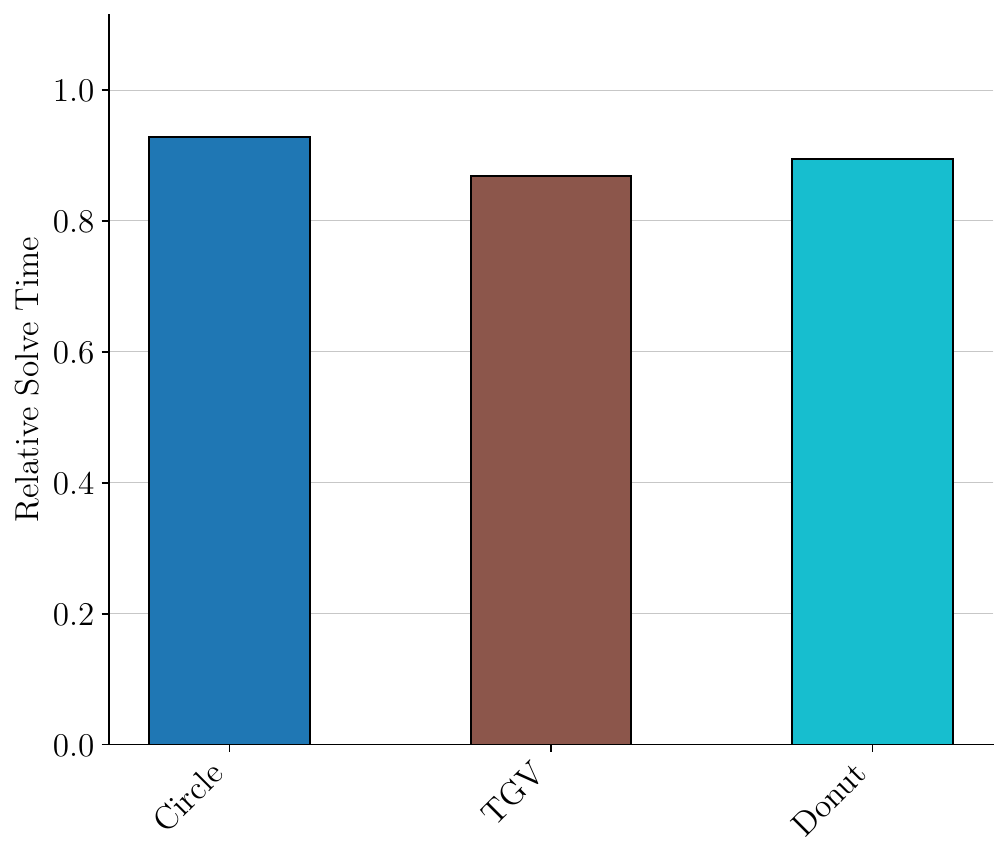}
        \caption{}
        \label{fig:solvetime.a}
    \end{subfigure}
    \hfill
    \begin{subfigure}[t]{0.50\linewidth} 
        \centering
        \includegraphics[width = \linewidth]{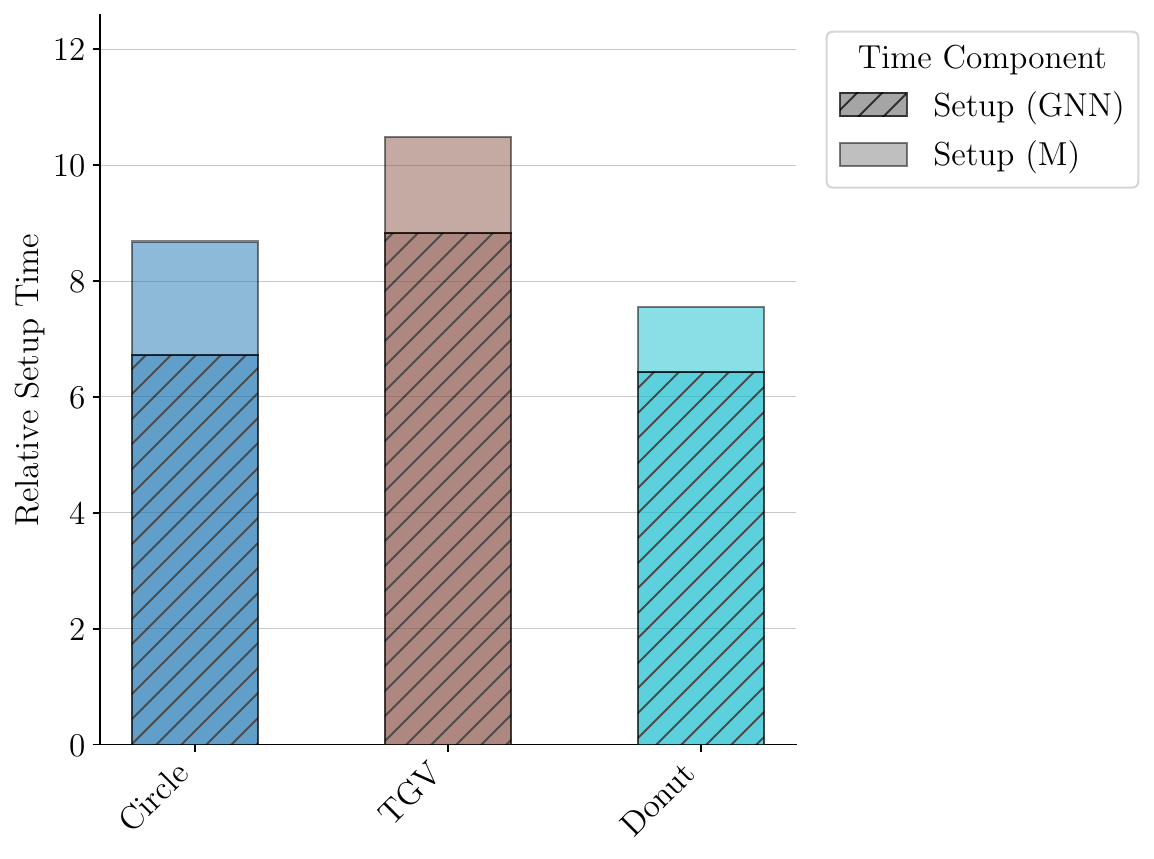}
        \caption{}
        \label{fig:setuptime.b}
    \end{subfigure}
    \caption{Mean GNN-based smoother (a) relative solve times and (b) relative setup times of 3 models normalized against the mean relaxed Jacobi solve time for 25 circle, TGV, and donut problems. Test dataset grids are $512\times 256$, $64^3$ and $128\times 64^2$, respectively.}
    \label{fig:solvetimewaterlily}
\end{figure}

An average solve time acceleration of $12\%$ is obtained for the unsteady flow test problems, as shown in Figure \ref{fig:solvetime.a}). We also note that all problems avoid divergence. This demonstrates that the GNN generalizes to problems considerably larger than the training dataset, across 2D and 3D grids and different types of flows. Notably, for an identical mesh size, the coefficient matrices $\mathbf{A}$ are equal between problems. The only mechanism to create different coefficient matrices is to increase the grid size, resulting in larger (but similar) graphs. Hence, the performance of the model is mainly limited by the complexity of the RHS vector and how well the pseudo-inverse matrices are constructed on unseen coarse levels, as larger problems require deeper hierarchies. 

Inferring the polynomial coefficients and constructing the pseudo-inverse matrices are the practical computational bottlenecks of the data-driven GNN-AMG solver. As depicted in Figure \ref{fig:setuptime.b}, the total setup time can take up to 10.5 times the relaxed Jacobi setup time. Provided that the acceleration is adequate, the time overhead can be compensated in a reasonable number of time steps. This demonstrates practical utility, especially for unsteady simulations that require $10^3$--$10^6$ time steps. Furthermore, in simple problems such as the structured meshes, the setup overhead can be virtually halved by using a GNN with 32 hidden dimensions, with almost no loss in performance (Appendix \ref{app:structured}). This demonstrates that simple meshes with regular algebraic patterns require a small learning capacity. 

\subsection{Unstructured meshes}
The majority of model training and testing is conducted on canonical CFD flows, such as Poiseuille, channel, convection-diffusion, and flat plate flows. These are laminar flows generated in 6 different mesh types. For 2D, the meshes are (i) structured, (ii) triangle Delaunay, (iii) triangle-quad Delaunay, (iv) triangle advancing front, (v) triangle-quad advancing front, and (vi) triangle advancing front with orthogonal smoothing. The three-dimensional problems use (i), (ii), and (iii). The main characteristics of the dataset are summarized in Table \ref{tab:dataset_summary} and examples of 2D grids are presented in Appendix \ref{app:refresco}, Figure \ref{fig:meshes}. The software ReFRESCO \citep{MARIN} has been used to generate the canonical flows. See more information about the dataset in Appendix \ref{app:refresco}.

Additionally, to test the model on problems that represent the physics of relevant aerodynamic flows, the AirfRANS dataset is considered. This dataset focuses on practical two-dimensional $k-\omega$ SST RANS simulations with different airfoil geometries. Not included in the canonical flows training dataset, the AirfRANS cases are used as a benchmark to test generalization. The grids have 280000 cells, on average.

The V-cycle computational graph for automatic differentiation has a significant memory footprint. Therefore, FP32 is selected to allow for larger mini-batch sizes during training. Accordingly, models are trained and tested in FP32 except for the AirfRANS evaluation, which uses FP64 because solution stagnation occurs when solving in single precision. 

\begin{table}[htbp]
    \centering
    \caption{Summary of canonical flows, geometries, and grid configurations used for training and testing the data-driven smoother.}
    \label{tab:dataset_summary}
    \begin{tabular}{@{} l c p{3.5cm} p{5cm} p{3.5cm} @{}}
        \toprule
        \textbf{Dataset} & \textbf{Dim.} & \textbf{Flow Types} & \textbf{Mesh Types} & \textbf{Total Problems} \\
        \midrule
        
        \multirow{6}{*}{Canonical CFD} & \multirow{6}{*}{2D} & \multirow{6}{3.5cm}{Poiseuille, channel, convection-diffusion, flat plate} 
        & Structured & \multirow{6}{3.5cm}{400 \newline (72 unique matrices)} \\
        & & & Triangle Delaunay & \\
        & & & Triangle-quad Delaunay & \\
        & & & Triangle advancing front & \\
        & & & Triangle-quad advancing front & \\
        & & & Triangle advancing front with orthogonal smoothing & \\
        
        \midrule
        
        \multirow{3}{*}{Canonical CFD} & \multirow{3}{*}{3D} & \multirow{3}{3.5cm}{Poiseuille, channel, convection-diffusion, flat plate} 
        & Structured & \multirow{3}{3.5cm}{600 \newline (14 unique matrices)} \\
        & & & Triangle Delaunay & \\
        & & & Triangle-quad Delaunay & \\
        
        \midrule
        
        AirfRANS & 2D & Aerodynamic airfoil flows (RANS with $k-\omega$ SST) & Structured meshes (including highly stretched inflation layers) & 1000 \newline (Tested in mini-batches of 100) \\
        
        \bottomrule
    \end{tabular}
\end{table}

\subsubsection{Generalization to unseen meshes}\label{sec:fulltrain}

To evaluate the generalization capabilities of the data-driven smoother on unstructured meshes, we first establish a baseline training setup using coarse grid resolutions. The training dataset consists of canonical CFD flows distributed as follows:

\begin{itemize}
    \item \textbf{2D Training dataset:} 400 problems featuring 72 unique coefficient matrices, with grid sizes up to 19,702 cells.

    \item \textbf{3D Training dataset:} 600 problems featuring 14 unique coefficient matrices, with grid sizes up to 15,313 cells.
\end{itemize}

While the coefficient matrices ($\mathbf{A}$) are shared among several cases, each of the 1,000 total problems possesses a unique source term vector ($\mathbf{f}$). All training is conducted in single precision (FP32), which halves the memory footprint while maintaining identical performance to FP64. Furthermore, to assess the robustness of our framework to training stochasticity, we train an ensemble of 10 identical models on this combined dataset, varying only the random seed used for weight initialization. 

During the testing phase, these 10 trained models are evaluated on unseen medium and fine-grid versions of the same canonical problems. To test the models' upscaling capabilities, the test grids are significantly larger than the training data: medium grids contain approximately 6 times as many cells as the largest training grids, while fine grids contain up to 20 times as many cells. Because these multiples are based on the largest training grids, the actual size difference between the average training problem and the test problems is even more pronounced.

\begin{figure}[h!]
    \centering
    \includegraphics[width=0.9\linewidth]{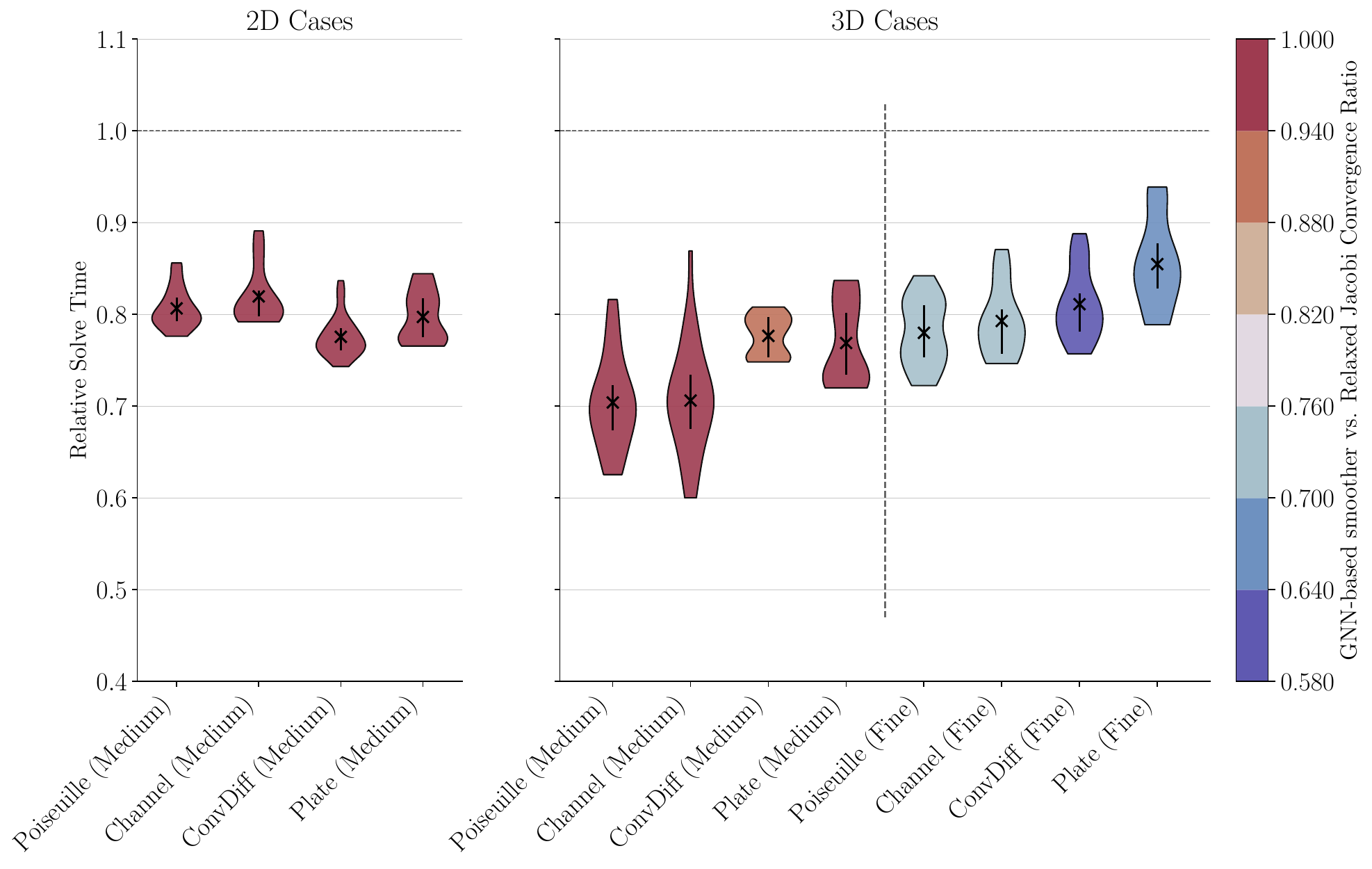}
    \caption{Relative solve time of GNN-based smoother against relaxed Jacobi for 10 models tested in canonical CFD problems. \textit{Medium} problems have approximately 5-6 more cells than the training dataset, while \textit{fine} around 20 times more. The cross mark is the mean of the set, and the thick line symbolizes the interquartile range, from the first percentile to the third.}
    \label{fig:relative_solve_mixed2d3d}
\end{figure}

The resulting average solver performance of these models is then evaluated relative to a baseline relaxed Jacobi smoother. Convergence ratios and average relative solve times per problem category are presented in Figure \ref{fig:relative_solve_mixed2d3d} and Table \ref{tab:unstruct_avg}.

The average solve time across the different models demonstrates that the data-driven smoother generalizes to larger unstructured meshes within the same problem type (Table \ref{tab:unstruct_avg}). Notably, some problems display large variation, such as medium Poiseuille and channel problems, reaching up to 66\% acceleration (see Figure \ref{fig:relative_solve_mixed2d3d}). Simpler problems, such as the medium 3D or the 2D ones, display, on average, adequate convergence ratios relative to relaxed Jacobi, above 96.1\%. However, the most complex grids, and more specifically convection-diffusion flows, exhibit a significant decrease in convergence. On average, 67.7\% convergence is achieved compared to relaxed Jacobi.

\begin{table}[h!]
\caption{Average acceleration and convergence success ratio of 10 trained models on unstructured data.}
\label{tab:unstruct_avg}
\begin{center}
\begin{tabular}{@{}lccc@{}}
    \toprule
    \textbf{Problems}             & 2D (Medium) & 3D (Medium) & 3D (Fine) \\ \midrule
    \textbf{Average acceleration (\%)} & 25      & 35.3     & 23.5   \\ 
    \textbf{Average convergence (\%)} & 97.9     &    96.1     & 67.7\\ \bottomrule
\end{tabular}
\end{center}
\end{table}

As previously shown on structured grids (Section \ref{sec:struct}, Figure \ref{fig:residual.a}), data diversity (especially regarding matrix coefficients) is essential for generalization. Consistent with those findings, we observed that training on a mixed dataset of two- and three-dimensional unstructured problems provided the most robust performance in terms of both acceleration and convergence compared to dimension-specific models. Furthermore, careful constraint of model complexity is required to prevent overfitting to the training graphs. As discussed in Section \ref{sec:3.1} regarding the polynomial degree, and detailed in the hyperparameter study in Appendix \ref{app:hyperparam_opt} (Table \ref{tab:size_gnn}), utilizing more complex models, such as third-degree polynomials, hidden dimensions larger than 64, or additional GNN layers, resulted in models that overfit the training data and performed considerably worse on unseen problems.

The large reduction in convergence is explained by observing at which hierarchy level the data-driven V-cycle starts to diverge. Accordingly, the residuals start to diverge mostly around the 3-4 AMG levels ($\sim3.7$ average), corresponding to meshes with $\sim2246$ cells. These mesh sizes are of the order of magnitude of the training dataset. However, the ones found during testing have considerably larger graph degrees, $\sim33.7$ on average. For a training graph of the same size, 3 to 6 edges are attached to each node, whereas in the evaluation the GNN must infer graphs with 34 edges per node. Therefore, the model is extrapolating and failing on 1/3 of the problems, as it is trying to generate polynomial coefficients for a graph configuration well outside the training range. This, however, does not prevent the GNN from constructing optimal pseudo-inverse smoothers for large meshes. In fact, many coarse vectors $\mathbf{f}^{(l)}$ lie within the training range and, although the graphs have larger degrees, the V-cycle with the data-driven smoother converges (faster than relaxed Jacobi). Hence, we observe that the main limitations of the model are:

\begin{enumerate}
    \item Inferring polynomial coefficients from graphs of similar size to the training data, but presenting a substantially higher graph degree (node connections).

    \item Applying the data-driven smoother on unseen Poisson problems with more complex flows (tougher $\mathbf{f}$), \textit{i.e.}, such as complex flow patterns with a wider range of scales.
\end{enumerate}

One approach to reduce these effects would be to introduce polyhedral meshes into the training dataset. By introducing polyhedral elements, the graph degrees would increase significantly (up to 16 or more). Thus, this could effectively increase the range of degrees the GNN observes without increasing the size of the dataset meshes.

\subsubsection{Generalization to new flows}\label{sec:halftrain}

\begin{figure}[h!]
    \centering
    \includegraphics[width=0.9\linewidth]{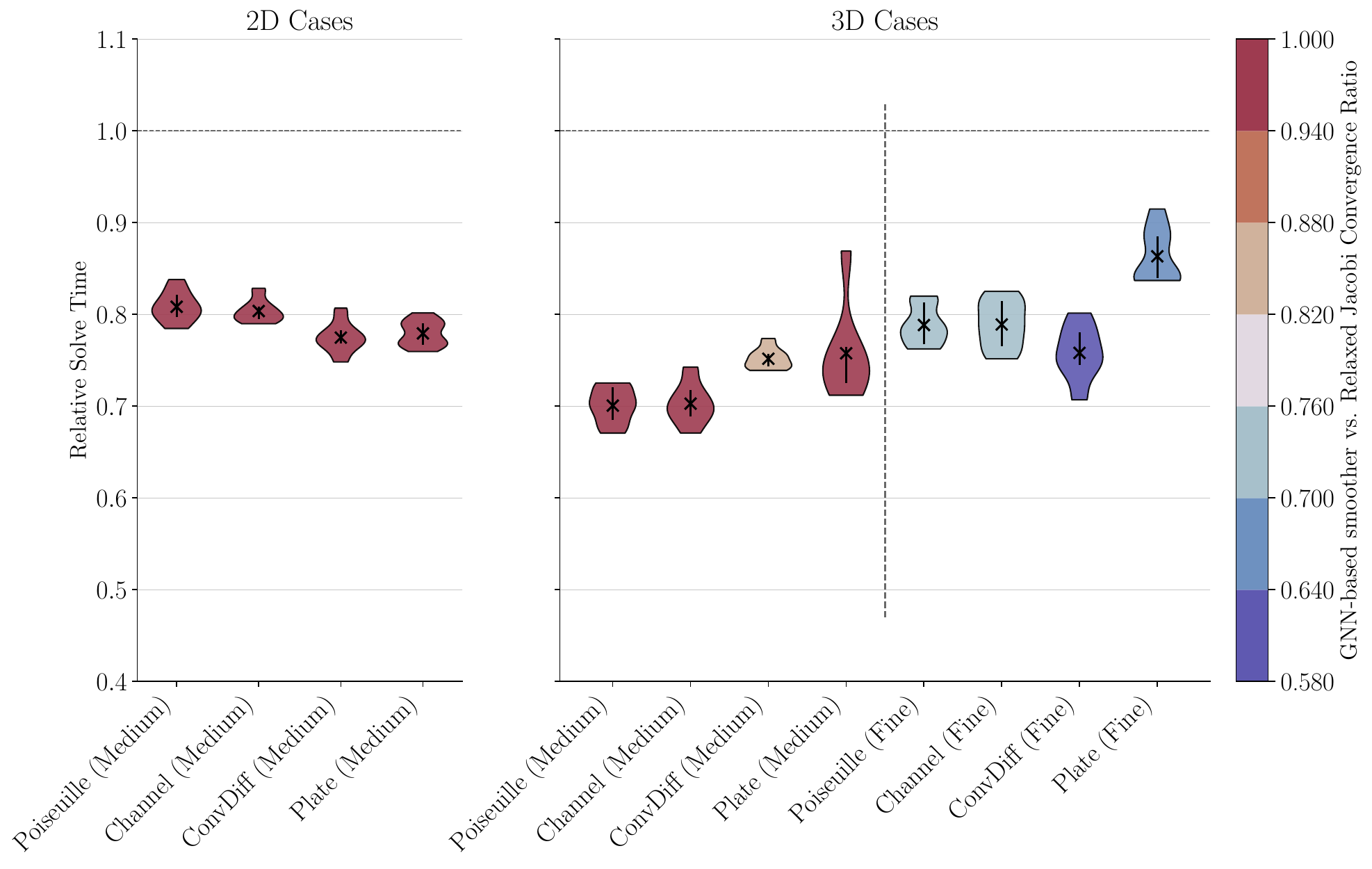}
    \caption{Relative solve time of data-driven smoother against relaxed Jacobi for 10 models tested in canonical CFD problems. Models are trained without channel and convection-diffusion problems.}
    \label{fig:relative_solve_noconvdiff}
\end{figure}

The model indirectly trains using information from the $\mathbf{f}$ vectors in the loss function. However, it mainly learns the algebraic features of the matrix, as $\mathbf{f}$ is not used in the GNN inference to predict optimal coefficients. Note that adding information of $f_i$ at each node attribute vector would imply evaluating the GNN at each time step, making the approach unfeasible. Thus, to address whether the model is able to generalize to unseen flows (unseen $\mathbf{f}$), we train 10 models exclusively on 300 (3D) and 200 (2D) Poiseuille and plate problems, respectively. We selected these two problem categories from the training dataset because they represent the boundaries of the solver's performance spectrum. Specifically, the Poiseuille cases provide a baseline for the simplest and fastest-solving flows, while the flat plate represents the slowest and most computationally demanding cases. These trained models are then assessed across all problem categories, with the relative solve times presented in Figure \ref{fig:relative_solve_noconvdiff}.  

The violin plot confirms two key aspects. First, when evaluated on channel and convection-diffusion flows, the models trained exclusively on Poiseuille and flat plate flows outperform the baseline models that have been trained on the full dataset. This demonstrates successful generalization to new flows with similar matrix structures (see Table \ref{tab:unstruct_avg_nochcd}). Second, restricting the training dataset exclusively to Poiseuille and flat plate flows actively reduces model overfitting. This dataset reduction discards hundreds of problem instances that simply reuse the exact same coefficient matrices ($\mathbf{A}$) with slightly different source terms ($\mathbf{f}$). Because the total number of unique matrices remains constant (72 for 2D, 14 for 3D), removing these repetitive problems increases the ratio of unique matrices to total training samples. Since the network's learning is driven primarily by $\mathbf{A}$ rather than $\mathbf{f}$, this concentrated dataset provides a more structurally varied training signal, preventing the network from over-memorizing duplicate matrices. This reduction in overfitting is reflected in a decrease in solve-time variability across all tested problems. Regarding convergence, both approaches achieve similar convergence rates, though the latter yields slightly lower values on medium 3D meshes, especially for convection-diffusion flows.

\begin{table}[h!]
\caption{Average acceleration and convergence success ratio of 10 trained models without channel and convection-diffusion on unstructured data.}
\label{tab:unstruct_avg_nochcd}
\begin{center}
\begin{tabular}{@{}lccc@{}}
    \toprule
    \textbf{Problems}             & 2D (Medium) & 3D (Medium) & 3D (Fine) \\ \midrule
    \textbf{Average acceleration (\%)} & 26.3    & 37.4      & 25      \\ 
    \textbf{Average convergence (\%)} & 99.7     &    93.8     & 68.1 \\ \bottomrule
\end{tabular}
\end{center}
\end{table}

\subsubsection{Performance in AirfRANS}\label{sec:airfrans}

\begin{figure}[h!]
    \centering
    \begin{subfigure}[t]{0.48\linewidth} 
        \centering
        \includegraphics[width = \linewidth]{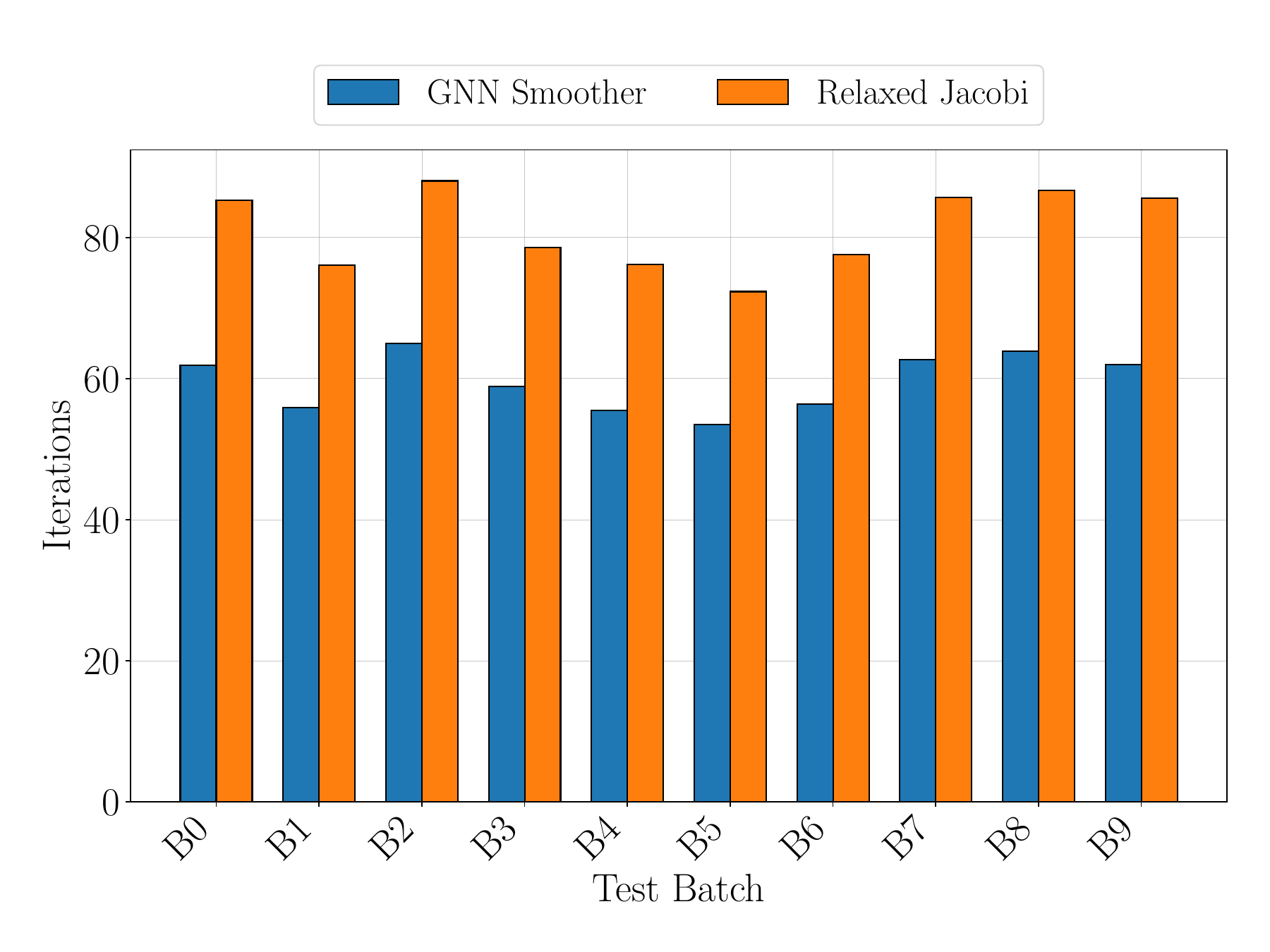}
        \caption{}
        \label{fig:airfrans.b}
    \end{subfigure}
    \hfill
    \begin{subfigure}[t]{0.48\linewidth} 
        \centering
        \includegraphics[width = \linewidth]{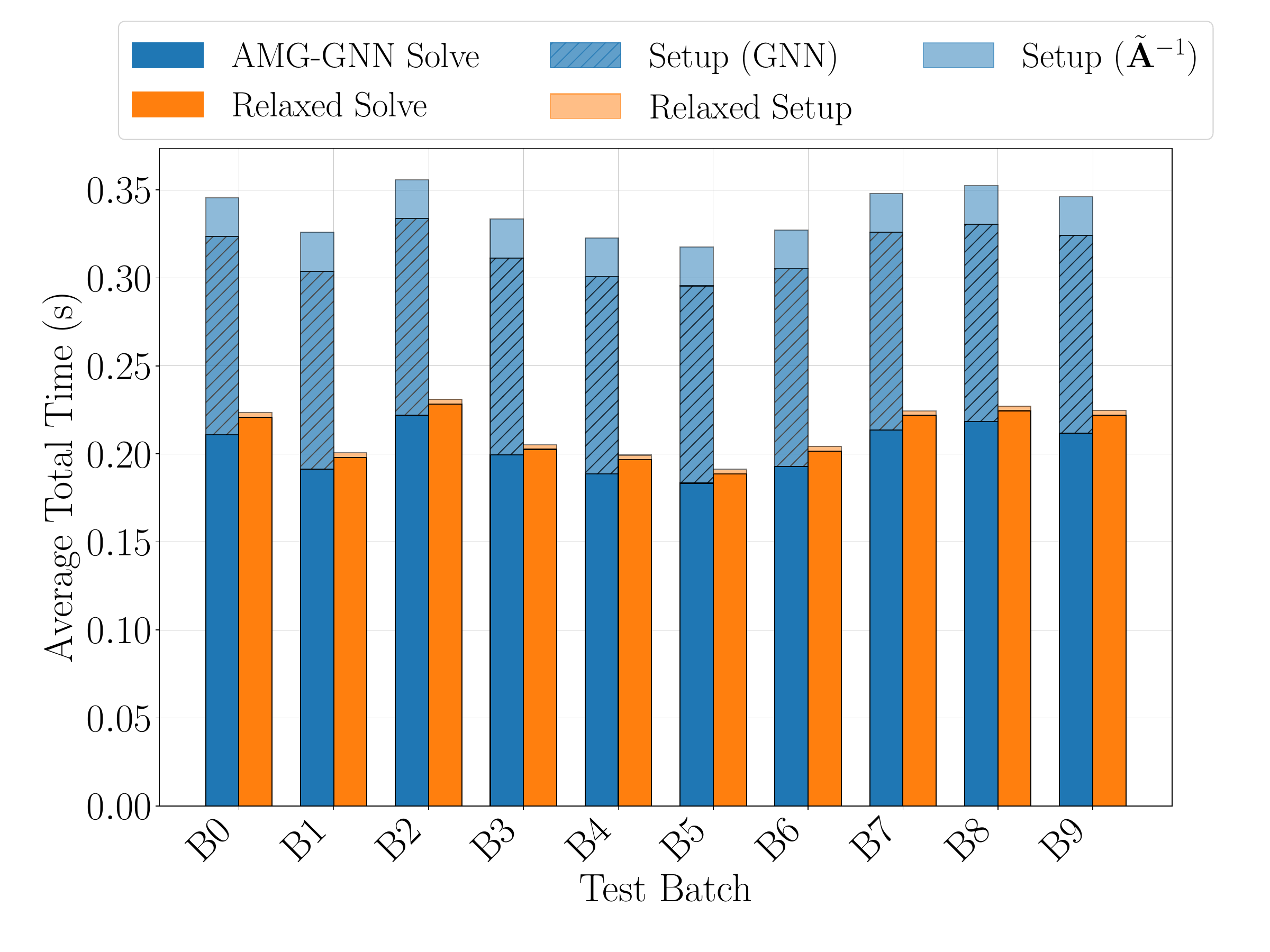}
        \caption{}
        \label{fig:airfrans.a}
    \end{subfigure}
    \caption{Performance of the best AMG-GNN model in the AirfRANS dataset versus the relaxed Jacobi baseline in terms of (a) average number of V-cycles per batch and (b) average total time per batch. Test batch size is 100.}
    \label{fig:airfrans}
\end{figure}

Finally, the best model trained on the complete dataset with all case categories (Section \ref{sec:fulltrain}) is tested on the AirfRANS dataset. The test dataset consists of 1,000 problems, evaluated in 10 mini-batches of 100 cases each. We present the results grouped by these mini-batches to preserve visibility into how performance scales across different mesh batches, while evaluating the individual setup and solve times naturally without performance variations caused by a single bulk calculation. Results for V-cycle iterations reduction are displayed in Figure \ref{fig:airfrans.b}. The model's performance demonstrates that it generalizes to unseen problems, including complex geometries and new flows, with 99.9\% convergence rates. Furthermore, it can predict optimal pseudo-inverse matrices for problems with highly stretched inflation layers, which have never been encountered during training. The model presents an approximate 27\% reduction in V-cycles. Note that the iteration reduction is the measure most closely related to the residual-reduction loss function used during training.

To show the practical utility of the model, Figure \ref{fig:airfrans.a} presents the total time corresponding to a time step solution with the pseudo-inverse prediction. The average solve time acceleration is 4\%. Compared to previous datasets, the performance of the model has diminished. Even though the reduction in number of iterations is large, using a sparse smoother instead of a diagonal one adds extra cost per iteration, hindering the solve time. Nonetheless, for this dataset, the reduction in solve time can be compensated in less than 30 time steps. 

The results demonstrate that models trained on flows without solid walls and on canonical flow cases can generalize to meshes with complex geometries, such as airfoils, and to meshes not seen during training, such as fine, inhomogeneous structured grids with inflation layers. Therefore, achieving a net 4\% acceleration opens several avenues for fine-tuning the dataset and the GNN architecture to achieve further speedups.

\section{Conclusions}\label{sec:5}

This study demonstrates that graph neural networks can effectively bridge the gap between fixed-parameter linear solvers and the complexity of unstructured mesh discretization. By substituting standard relaxation parameters with a GNN-based sparse pseudo-inverse smoother $\tilde{\mathbf{A}}^{-1}$, substantial reductions in both iteration counts and total wall-clock time were achieved while strictly maintaining the underlying linearity of the solver. The modified GCIN architecture successfully learns algebraic patterns that generalize beyond specific flow configurations, enabling models trained on small synthetic cases to generalize to complex aerodynamic flows.

The numerical experiments demonstrate robust generalization across scaling factors, particularly on structured meshes where the GNN is trained on small $32^2$ and $16^3$ synthetic problems and successfully applied to grids up to 128 times larger. Despite this significant discrepancy between training and test resolutions, the model achieves an average solve-time acceleration of 12\% for unsteady flow benchmarks, such as the Taylor--Green vortex and flow past a donut. 

Transitioning to unstructured meshes, the framework is validated using canonical CFD flows across diverse 2D and 3D mesh topologies. Even when trained on coarse data, the enhanced smoother generalizes to fine, unstructured meshes with 20 times more cells, delivering average solve-time acceleration of 25\% in 2D and 23.5\% in 3D. These tests confirm that the GCIN architecture effectively extracts generalized algebraic features from varying connectivity patterns rather than overfitting to specific flow properties. 

The model's practical utility is confirmed using the industry-relevant AirfRANS dataset, which introduces complex geometries, turbulence models, and highly stretched inflation layers never encountered during training. In this challenging benchmark, the best data-driven smoother maintains a 99.9\% convergence rate while achieving a significant 27\% reduction in V-cycle iterations. Although the higher per-iteration cost of a sparse smoother limits the total wall-clock speedup to 4\% for this specific dataset, the results establish that the learned algebraic patterns are highly scalable and effective for practical aerodynamic simulations.

Finally, there are current limitations in the present model. First, the setup phase of the data-driven smoother is non-negligible but compensated by the solve-time speed-up when several evaluations are required (e.g., unsteady flows). Also, while the proposed data-driven smoother provides a significant speedup in solve time compared to other approaches in the literature, a slight deviation in the optimal coefficients can interfere with convergence. Still, we show that adding algebraically similar linear problems to the training dataset mitigates this issue. Last, the present work uniquely assesses the model's generalization to unseen, industry-relevant flows using unstructured meshes.

\section*{Acknowledgments}

BF acknowledges the financial support of the Dutch Research Council (NWO) Veni Talent Programme for the FastFlows project (file number 21886) under the grant ID \href{https://doi.org/10.61686/TPHBS84265}{doi.org/10.61686/TPHBS84265}.

\section*{CRediT author statement}

\begin{itemize}
    \item \textbf{Eric Chillón:} Data curation, Formal analysis, Investigation, Methodology, Software, Validation, Visualization, Writing – original draft, Writing – editing.

    \item \textbf{Artur K. Lidtke:} Conceptualization, Data curation, Supervision, Writing – review

    \item \textbf{Nguyen Anh Khoa Doan:} Conceptualization, Resources, Supervision, Writing – review \& editing

    \item \textbf{Bernat Font:} Conceptualization, Funding acquisition, Project administration, Supervision, Writing – original draft, Writing – review
\end{itemize}

\bibliographystyle{etc/apalike-refs}
\bibliography{main}

\appendix

\section{Model architecture}

Here we detail the GNN architecture (Appendix \ref{app:hyperparam_arch}), which is used throughout all experiments. The architecture size has been selected based on a preliminary hyperparameter study presented in Appendix \ref{app:hyperparam_opt}. Some architectural choices, such as normalization or concatenation, have been chosen based on trial and error.

\subsection{Architectural hyperparameters}\label{app:hyperparam_arch}

To predict the optimal coefficients for the sparse pseudo-inverse smoother, we employ a modified GCIN architecture. This model is designed to extract expressive features from the algebraic structure of the sparse matrix $\mathbf{A}$ across different hierarchy levels. The architecture consists of three primary components: feature initialization, a series of message-passing layers, and a global readout mechanism followed by a predictive MLP.

The sparse coefficient matrix $\mathbf{A}$ is mapped to a graph $G = (V, E)$, where nodes $v_i \in V$ represent grid cells and directed edges $e_{ij} \in E$ represent their algebraic connections. Each node is initialized with a two-dimensional feature vector $\mathbf{x}_{v_i}^{(0)} = [a_{ii}, deg(v_i)]^T$, containing the matrix diagonal coefficient and the node degree. The node degree informs the network about the mesh dimensionality (2D vs. 3D) and element types. Edges are assigned scalar attributes $w_{ij}$ equal to the off-diagonal coefficients $a_{ij}$.

The core of the feature extractor consists of four sequential GCIN layers. The GCIN combines the efficiency of GCNs with the expressiveness of GINs. The forward pass for a layer $l$ is defined as

$$\mathbf{x}_{v_i}^{(l)} = \text{MLP}^{(l)} \left( \sum_{v_j \in N(v_i)\cup v_i} w_{ij}\mathbf{x}_{v_j}^{(l-1)}\right).$$

To handle the grid size variance between problems, the model utilizes instance normalization (InstanceNorm) after each convolution instead of standard batch normalization. This ensures the network normalizes features per-graph. Following, a ReLU activation function is applied after normalization in each layer.

The architecture employs a jumping knowledge (JK)-like strategy \citep{xu2018representationlearninggraphsjumping} to aggregate information from all $L$ layers, capturing both local and more distant neighborhood patterns. For each layer $l$, the node representations are processed through both global mean pooling and global max pooling to retain both average structural trends and extreme algebraic variations. Next, the pooled results from each layer are concatenated and passed through a linear layer. These representations are then concatenated into a final graph embedding vector of size $128 \times L$. This global embedding is passed through a layer normalization step and a sigmoid non-linearity before the final three-layer MLP with ReLU activations. The sigmoid proved to converge faster and perform better than other activation functions such as ReLU or tanh. Finally, the MLP maps the features to 5 output values: 3 coefficients for the diagonal polynomial $f_d$ and 2 for the off-diagonal polynomial $f_o$ (where $f_o(0)=0$ is strictly enforced).

The total number of trainable parameters is 55045. The following table summarizes the hyperparameter configuration for the GNN architecture used throughout the experiments.

\begin{table}[h!]
\centering
\caption{Hyperparameters for the GNN model architecture.}
\label{tab:gnn_hyperparams}
\begin{tabular}{@{}ll@{}}
\toprule
\textbf{Hyperparameter} & \textbf{Value} \\ \midrule
Number of GCIN layers ($L$) & 4  \\
Input node feature dimension & 2  \\
Hidden layer dimension & 64  \\
Global pooling strategy & Concatenated Mean and Max  \\
Readout aggregation & Jumping Knowledge (Concatenation)  \\
MLP number of layers & 2 \\
MLP hidden dimension & 128  \\
Output dimension (coefficients) & 5 ($3$ for $f_d$, $2$ for $f_o$)  \\
Normalization (Convolution) & Instance Normalization \\
Normalization (Readout) & Layer Normalization \\
Activation function & ReLU \\ \bottomrule
\end{tabular}
\end{table}


\subsection{Hyperparameter tuning}\label{app:hyperparam_opt}

We select an optimal GNN size based on a reduced 2D and 3D unstructured dataset during an initial assessment (up to 15000 cells). Performance is evaluated on purely 2D unstructured problems (around 80000 cells). The two hyperparameters that are evaluated are the GCIN hidden layer dimension and the number of GCIN layers. We do not test different MLP layer dimensions since we construct them based on the GCIN output. Other architectural decisions have been decided based on trial and error, starting from the original architecture foundation \citep{HaifengZou2024}.

Results are presented in Table \ref{tab:size_gnn}. Except for the reduced GCIN layer configuration for plate flows, the $4\times64$ configurations prove to be the best for unstructured problems. The range of features that the model needs to learn is better captured on a medium to large-sized GCIN. Going to larger hidden dimensions provides little to no performance and a noticeable jump in GNN inference time. Moreover, there is a limit at which obtaining information from farther neighbors (adding more GCIN layers) does not add any advantage to the network. 

\begin{table}[h!]
\centering
\caption{Relative solve times for several GCIN configurations.}
\label{tab:size_gnn}
\begin{tabular}{@{}cccccc@{}}
\toprule
\multirow{2}{*}{\textbf{Hyperparameter}} & \multirow{2}{*}{\textbf{Value}} & \multicolumn{4}{c}{\textbf{Relative solve time (\%)}} \\
 &  & Poiseuille & Channel & Convection-Diffusion & Plate \\ \midrule
\multirow{3}{*}{\textbf{Hidden layer dimension}} & \textbf{64} & \textbf{83.8} & \textbf{84.6} & \textbf{80.0} & \textbf{83.8} \\
 & 32 & 88.0 & 87.7 & 84.4 & 85.0 \\
 & 16 & 90.6 & 88.8 & 87.5 & 86.7 \\ \midrule
\multirow{3}{*}{\textbf{GCIN Layers}} & 6 & 85.3 & 84.8 & 80.3 & 83.9 \\
 & \textbf{4} & \textbf{83.8} & \textbf{84.6} & \textbf{80.0} & 83.8 \\
 & 2 & 84.6 & 92.9 & 82.0 & \textbf{81.4} \\ \bottomrule
\end{tabular}
\end{table}



\section{Experiment details}

In this section, we provide experiment details for structured and unstructured datasets. Firstly, we provide training details and a hyperparameter table in Appendix \ref{app:train_procedure}, followed by specific testing details in Appendix \ref{app:test_procedure}. Trainings and tests are carried out in an NVIDIA Quadro RTX 8000 (48GB)

\subsection{Training details} \label{app:train_procedure} \label{app:hyperpara_train}

\begin{figure}[h!]
    \centering
    \includegraphics[width=0.6\linewidth, trim={0pt 0pt 0pt 36pt}, clip]{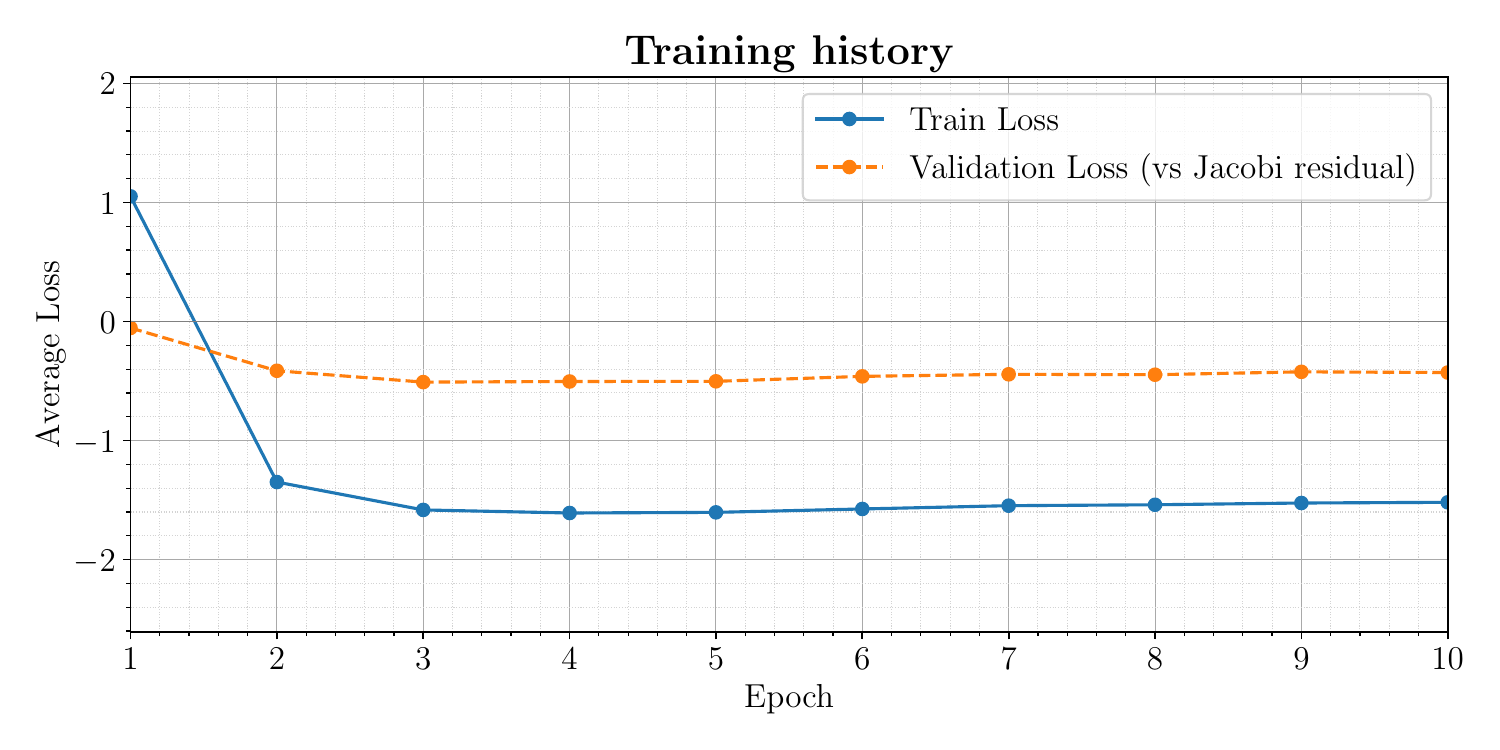}
    \caption{History of train and validation loss for the structured dataset.}
    \label{fig:app_struct_loss}
\end{figure}

\begin{figure}[h!]
    \centering
    \includegraphics[width=0.6\linewidth, trim={0pt 0pt 0pt 47.5pt}, clip]{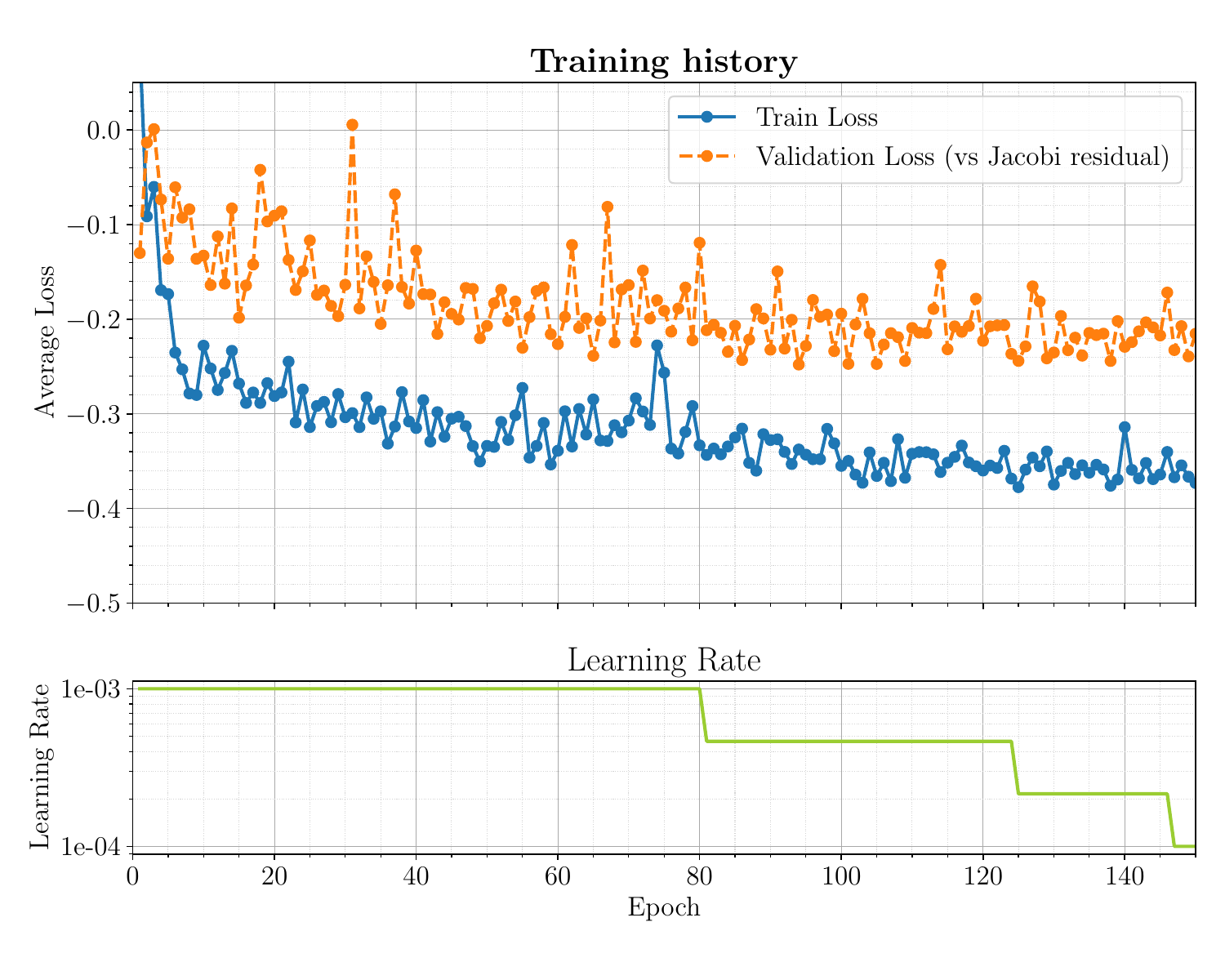}
    \caption{History of train and validation loss for the unstructured dataset. The learning rate is displayed below to show the effect of the scheduler.}
    \label{fig:app_unstruct_loss}
\end{figure}

To keep the training simple, we maintain the GNN model size for structured and unstructured meshes. We adapt, however, the mini-batch size and the training epochs for each type of dataset. Training hyperparameters are presented in Table \ref{tab:train_hyper}. We do this to adjust the training to the dataset characteristics. Regarding the mini-batch size, structured problems have a smaller memory footprint, allowing for larger mini-batch sizes without stability issues and thus faster training (unstructured datasets can sporadically fill up the 48GB of GPU memory, crashing the training). Concerning the training epochs, structured datasets contain significantly less variety in terms of $\mathbf{A}$ matrices. Hence, the most optimal data-driven smoother is learned extremely fast, within 2-5 epochs (Figure \ref{fig:app_struct_loss}). This contrasts with the unstructured dataset, generally requiring around 80-100 epochs (Figure \ref{fig:app_unstruct_loss}).

Training for structured datasets takes approximately 30 minutes. Conversely, the process for unstructured datasets requires up to 4 hours. Despite the reduced number of training epochs, the sequential V-cycle inference is the largest performance bottleneck regarding wall-clock training times.

Finally, as an early-stopping criterion, we use the best validation model. When left unattended for many epochs, the GNN overfits the training data. Storing the model that achieves the lowest validation loss provides the best generalization performance when testing on unseen datasets, especially for structured meshes.

\begin{table}[h!]
\centering
\caption{Hyperparameters for data-driven AMG smoother training}\label{tab:train_hyper}
\begin{tabular}{@{}ll@{}}
\toprule
\textbf{Hyperparameter} & \textbf{Value / Configuration} \\ \midrule
$\nu_1$ (Pre-smoothing steps) & 2 \\
$\nu_2$ (Post-smoothing steps) & 2 \\
Max AMG hierarchy levels & 10 \\
Optimizer & Adam \\
Learning rate ($\eta$) & $10^{-3}$\\
Optimizer scheduler & ReduceLROnPlateau \\
Mini-batch size & 10 (Structured) / 5 (Unstructured) \\
Training epochs & 30 (Structured) / 150 (Unstructured) \\
Gradient clipping (Max norm) & 1.0 \\
\bottomrule
\end{tabular}
\end{table}

\subsection{Testing details}\label{app:test_procedure}

The testing is carried out following the training process, but without mini-batching the datasets. The problems are tested individually. By evaluating the Poisson systems one at a time, three problem-specific time measures are achieved: the GNN inference, the pseudo-inverse construction, and the solve time to reach a certain tolerance. When evaluating non-trainable smoothers such as the relaxed Jacobi, only the pseudo-inverse construction and solve times are determined.

Generally, when running a program for the first time, the code execution is slower than in subsequent runs. This is due to a \textit{cold start} and the metrics should be removed. To this end, we perform multiple calculations for each time metric, and we remove the initial and average the remaining ones. This also reduces the measured time uncertainty, as each run is slightly different from the others (\textit{e.g.}, caused by system noise).

The speed of this testing approach largely depends on the problems that are being evaluated (size, dimension, complexity). For instance, the unsteady flow structured datasets (25 problems) take around 15 seconds for circle, 40 seconds for TGV, and 50 seconds for donut. On the other hand, the largest meshes (200 problems) on the unstructured dataset take approximately 12 minutes on average for each problem type.

\subsection{Optimizing GNN inference}\label{app:structured}

\begin{figure}[h!]
    \centering
    \begin{subfigure}[t]{0.46\linewidth} 
        \centering
        \includegraphics[width = \linewidth]{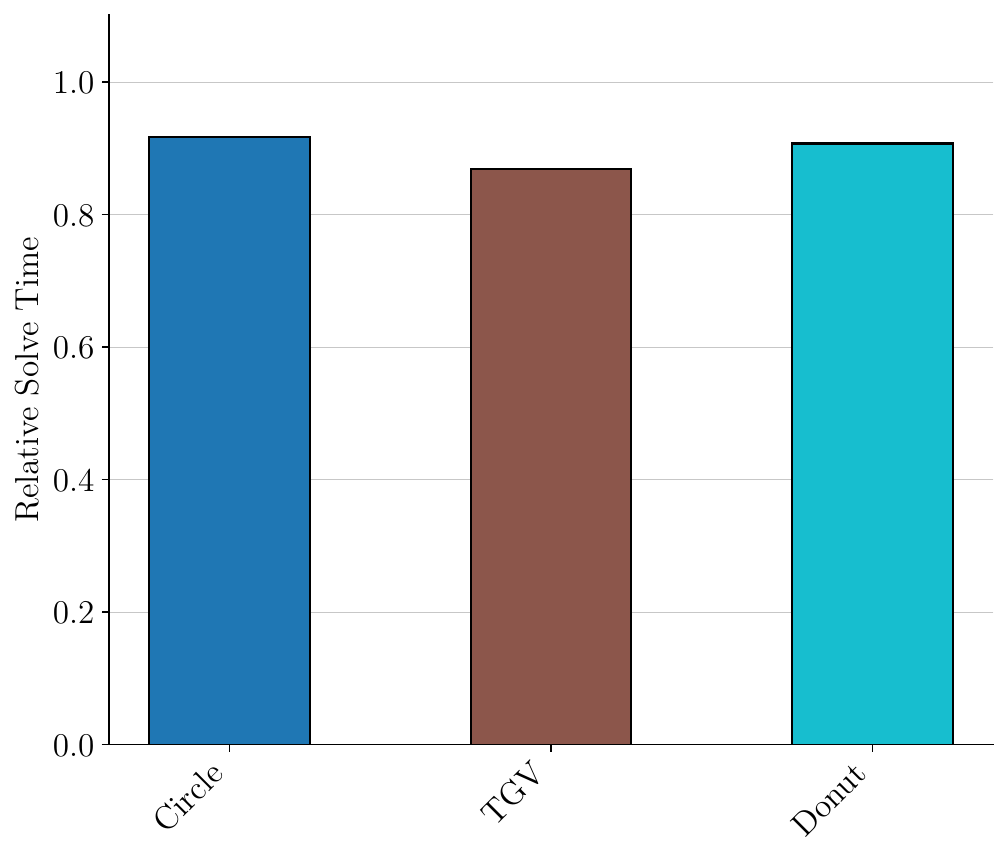}
        \caption{}
        \label{fig:solvetimea.a}
    \end{subfigure}
    \hfill
    \begin{subfigure}[t]{0.50\linewidth} 
        \centering
        \includegraphics[width = \linewidth]{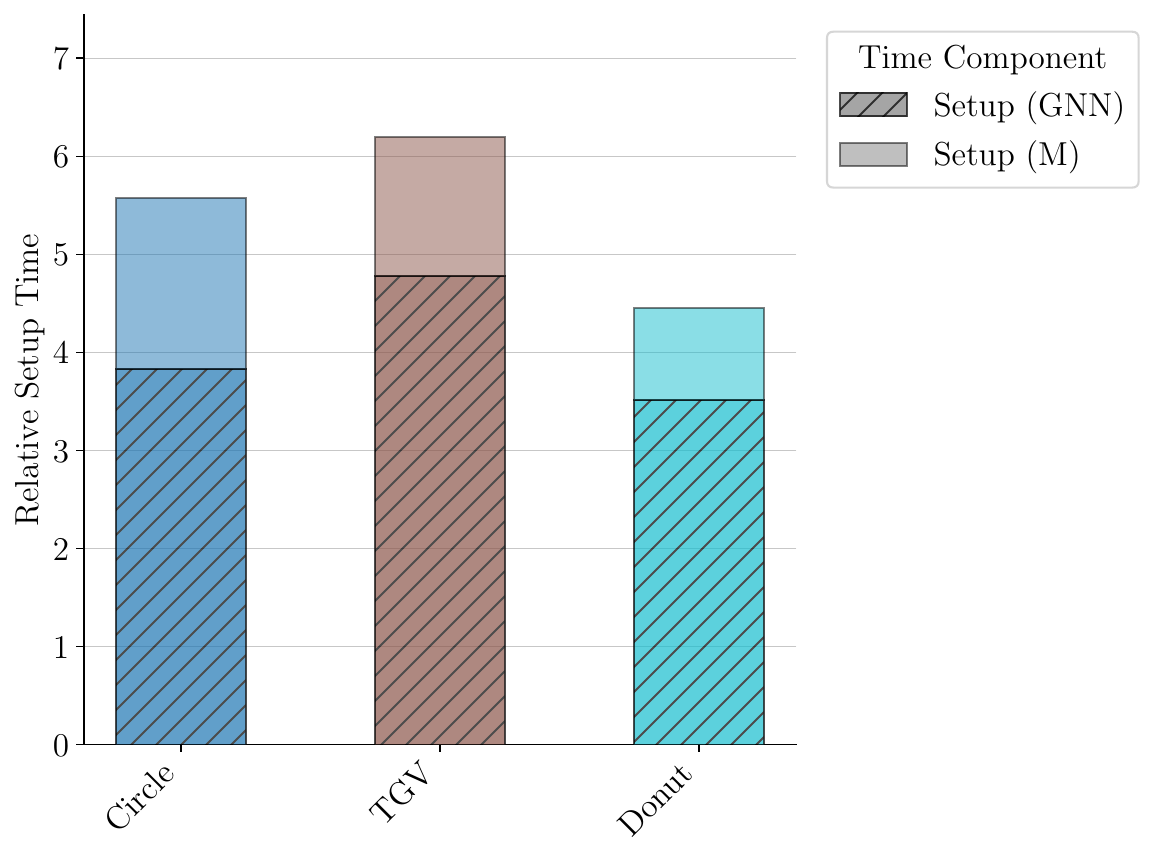}
        \caption{}
        \label{fig:setuptimea.b}
    \end{subfigure}
    \caption{Mean GNN-based smoother (a) relative solve times and (b) relative setup times of 3 trained models normalized against the relaxed Jacobi solve time for 25 circle, TGV and donut problems. The GCIN hidden dimension is 32. Test dataset grids are $512\times 256$, $64^3$ and $128\times 64^2$, respectively.}
    \label{fig:app_unsteady}
\end{figure}

The GNN inference is a computational bottleneck, with the forward pass time being highly dependent on the GNN block size. Simpler problems, such as those based on structured meshes, allow for reduced node dimensions (GCIN hidden dimension) due to the more regular algebraic patterns. Thus, on such problems, the hidden dimension size can be halved. The outcome is near-identical performance and a significant time and memory reduction. Relative solve and setup times for unsteady flow problems (circle, TGV and donut) for three 32-hidden dimension GCIN models are presented in Figure \ref{fig:app_unsteady}.

Although the data-driven smoother is aimed at unsteady meshes, this approach is also valid for structured heterogeneous meshes. Generally, these meshes contain large stretching such as inflation layers. Therefore, the classic geometric multigrid becomes harder to implement and AMG is usually chosen. Results throughout this study demonstrate practical utility in structured heterogeneous meshes such as those in the AirfRANS dataset. Hence, an efficient low-size GNN for structured-only meshes could be a viable approach. 

\section{Canonical flows details}\label{app:refresco}

To generate unstructured data, the ReFRESCO software is used. ReFRESCO is a multi-phase viscous flow solver focused on maritime applications under active development by the Maritime Research Institute Netherlands \citep{MARIN}. The main problems that comprise the unstructured dataset are Poiseuille, channel, Convection-Diffusion (ConvDiff), and plate flows. They are common CFD problems that allow for evaluating the GNN model outside synthetic benchmark problems for AMG and show performance towards a more applicable end. 

\begin{figure}[h!]
    \centering
    \begin{subfigure}[t]{0.32\linewidth}
        \centering
        \includegraphics[width = \linewidth]{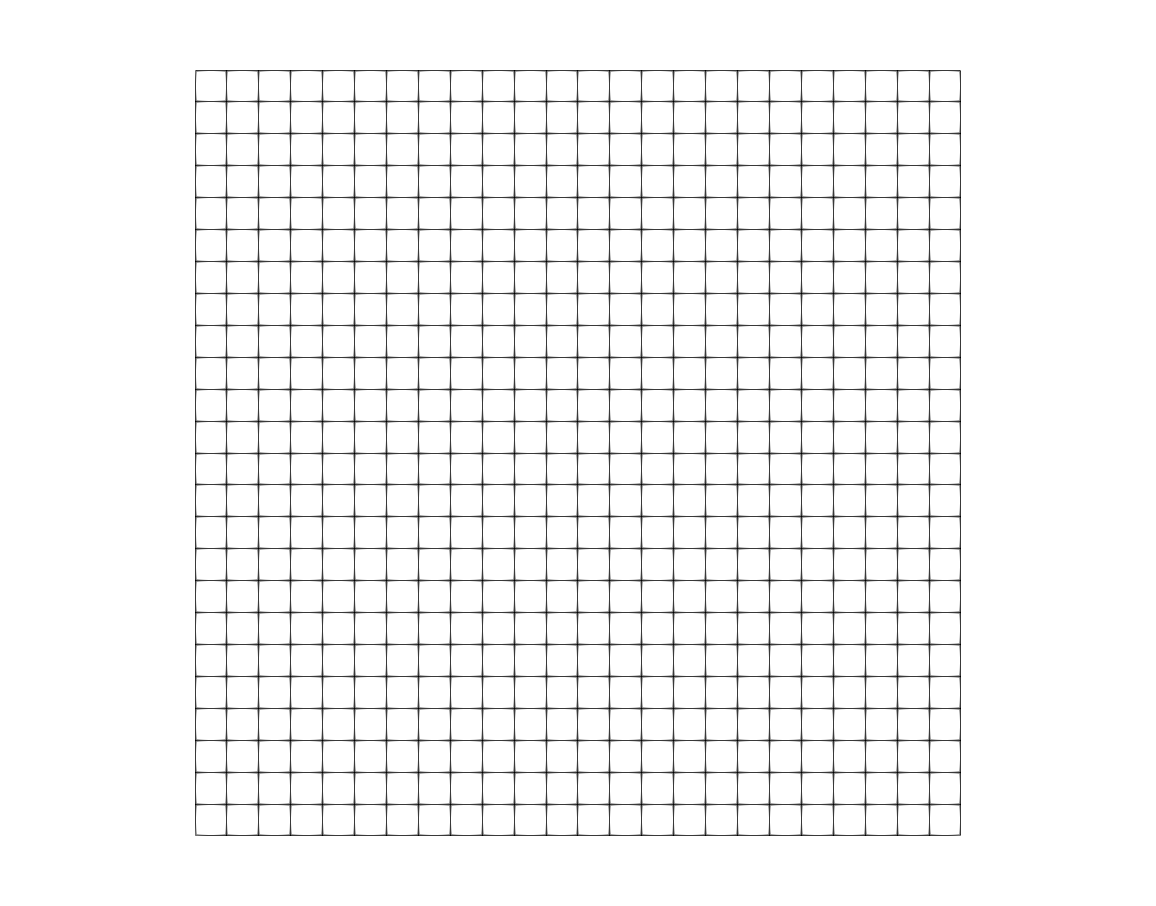}
        \caption{}
        \label{fig:5.1.2.meshes_1.a1}
    \end{subfigure}
    \hfill
    \begin{subfigure}[t]{0.32\linewidth}
        \centering
        \includegraphics[width = \linewidth]{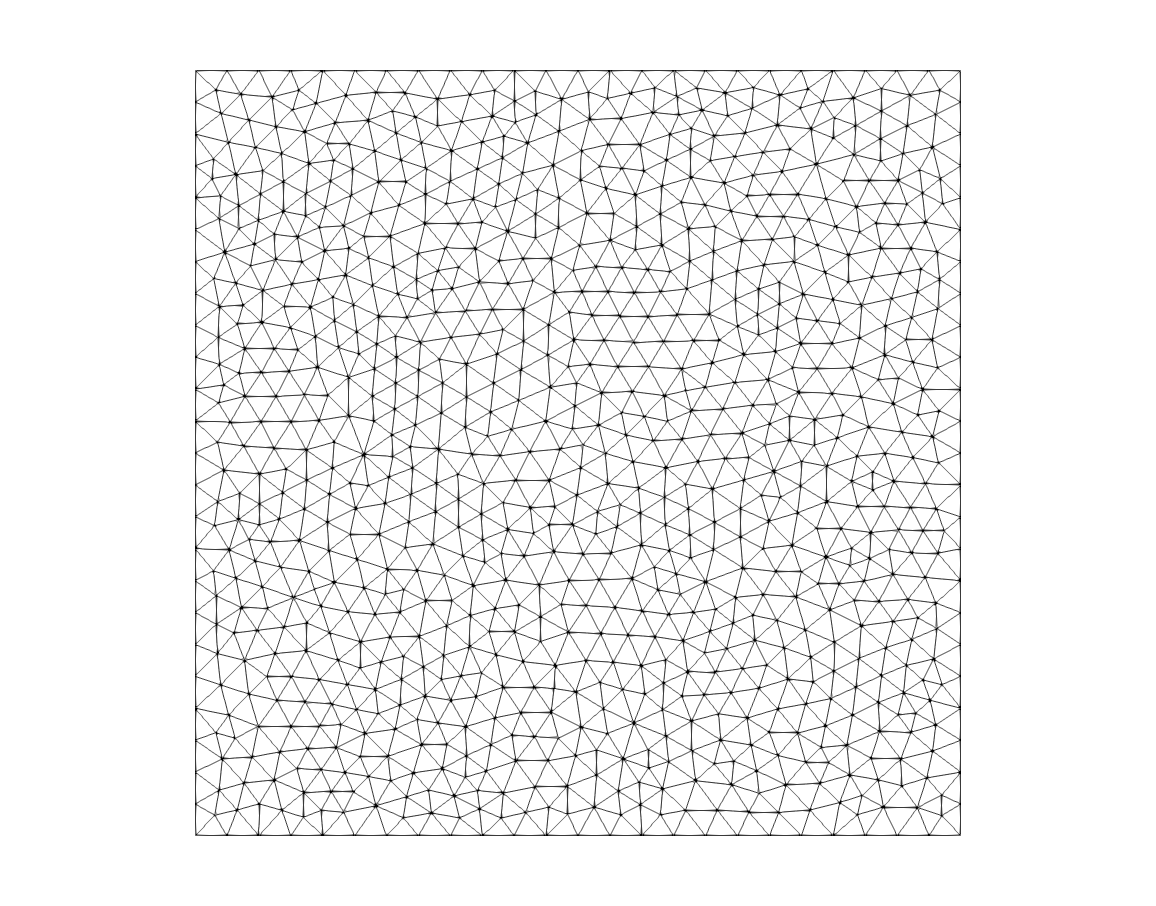}
        \caption{}
        \label{fig:5.1.2.meshes_1.b2}
    \end{subfigure}
    \hfill
    \begin{subfigure}[t]{0.32\linewidth} 
        \centering
        \includegraphics[width = \linewidth]{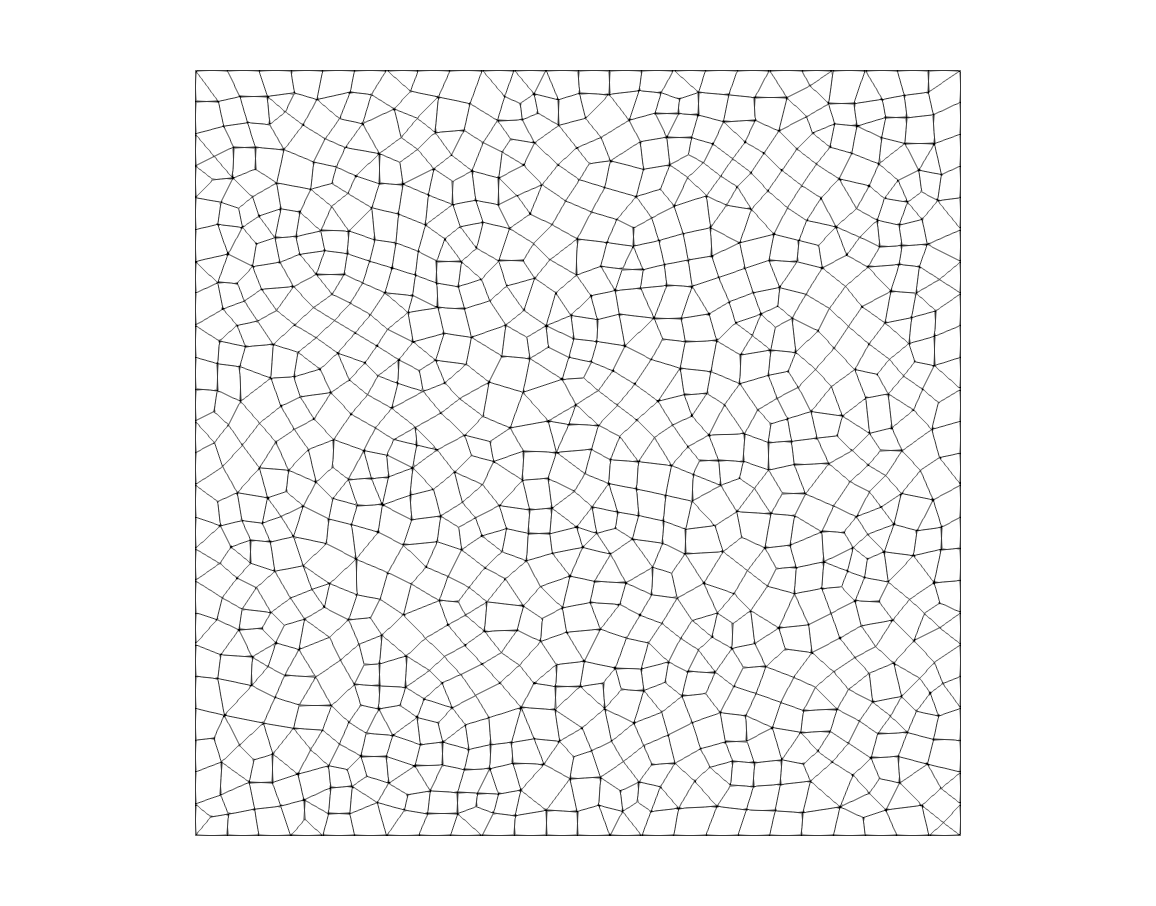}
        \caption{}
        \label{fig:5.1.2.meshes_1.c2}
    \end{subfigure}
    \begin{subfigure}[t]{0.43\linewidth}
        \centering
        \includegraphics[width = \linewidth]{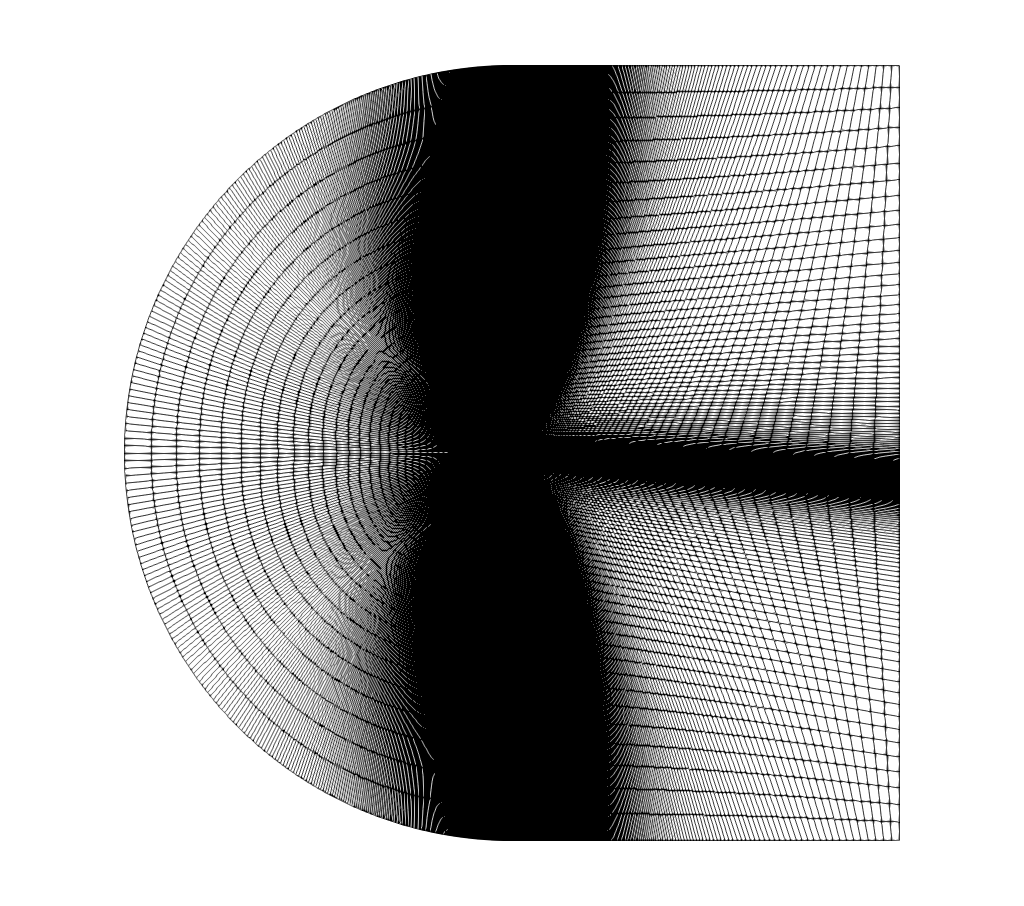}
        \caption{}
        \label{fig:5.1.2.meshes_2.a}
    \end{subfigure}
    \begin{subfigure}[t]{0.40\linewidth}
        \centering
        \includegraphics[width = \linewidth]{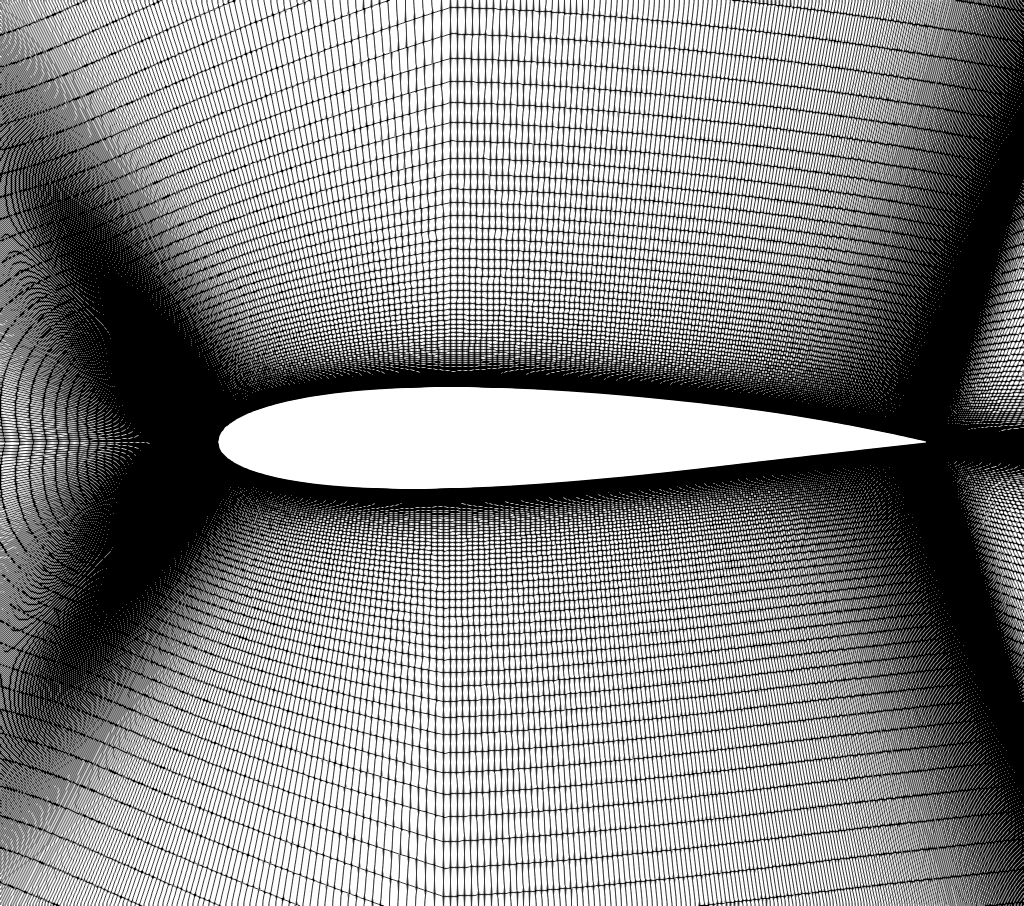}
        \caption{}
        \label{fig:5.1.2.meshes_2.b}
    \end{subfigure}
    \hfill
    \caption{Example of meshes used throughout the study are (a) structured, (b) triangle Delaunay, (c) quad-dominant, (d) AirfRANS far view and (e) AirfRANS zoomed view.}
    \label{fig:meshes}
\end{figure}

\begin{figure}[h!]
    \centering
    \begin{subfigure}[t]{0.61\linewidth} 
        \centering
        \includegraphics[width = \linewidth]{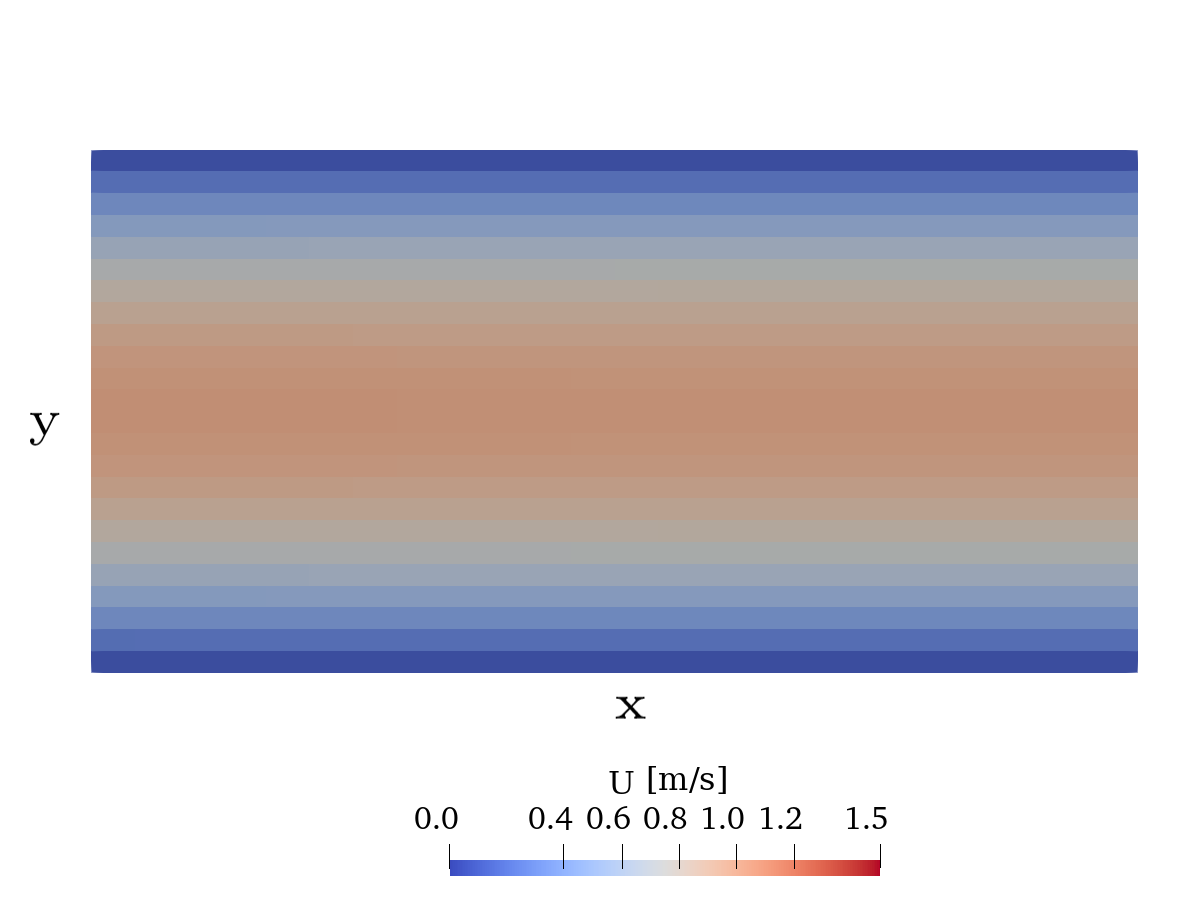}
        \caption{}
        \label{fig:5.1.2.schemes_poicha.a}
    \end{subfigure}
    \hfill
    \begin{subfigure}[t]{0.38\linewidth} 
        \centering
        \includegraphics[width = \linewidth]{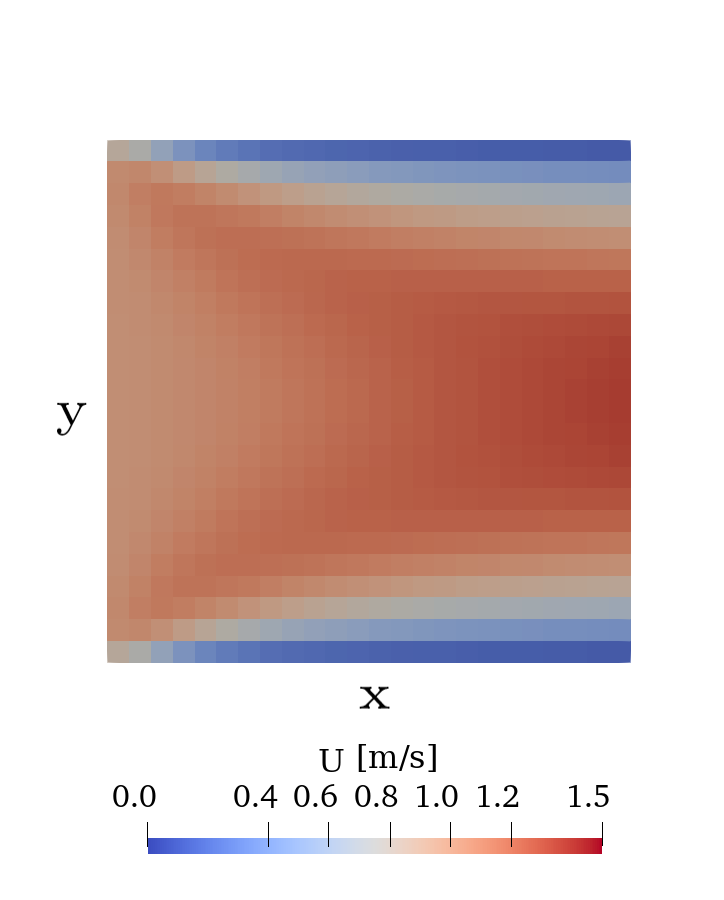}
        \caption{}
        \label{fig:5.1.2.schemes_poicha.b}
    \end{subfigure}
    \hfill
    \begin{subfigure}[t]{0.48\linewidth} 
        \centering
        \includegraphics[width = \linewidth]{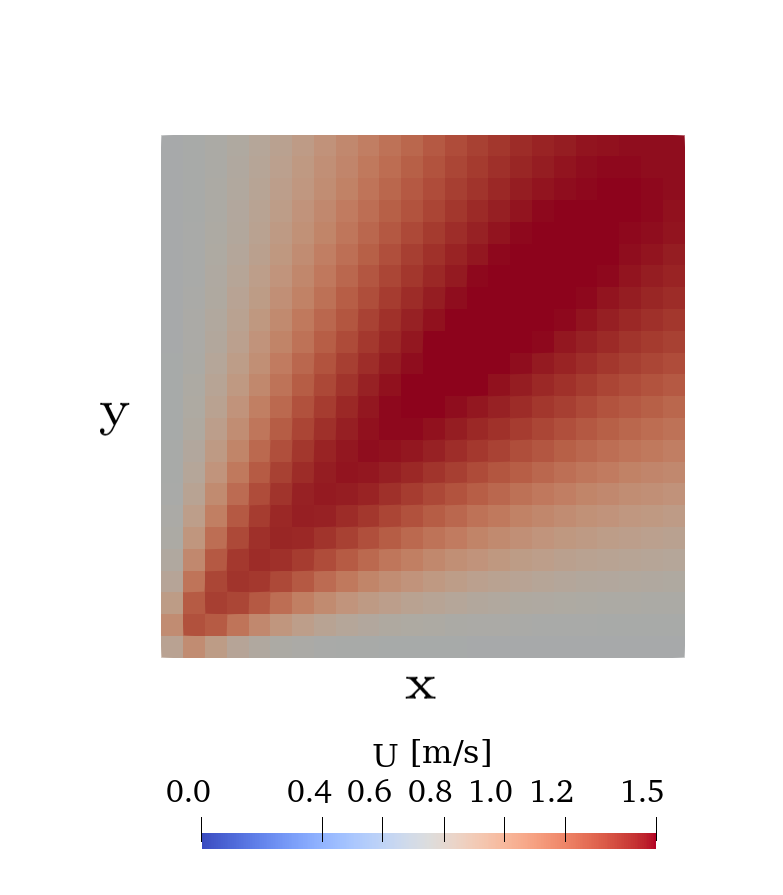}
        \caption{}
        \label{fig:5.1.2.schemes_cdplate.a}
    \end{subfigure}
    \hfill
    \begin{subfigure}[t]{0.46\linewidth} 
        \centering
        \includegraphics[width = \linewidth]{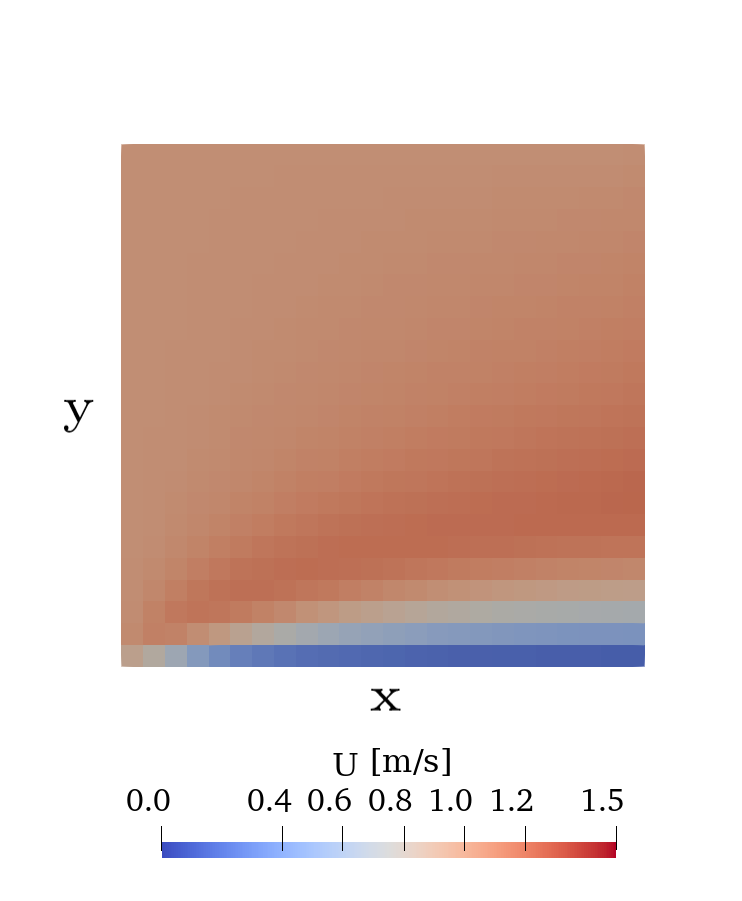}
        \caption{}
        \label{fig:5.1.2.schemes_cdplate.b}
    \end{subfigure}
    \caption{Canonical flow fields in structured meshes for (a) Poiseuille, (b) channel, (c) convection-diffusion and (d) flat plate flows.}
    \label{fig:canonicalCFD}
\end{figure}

All problems presented in Figure \ref{fig:canonicalCFD} are generated under laminar flow conditions. The setup and boundary conditions for each case are presented in Tables \ref{tab:setupcfd} and \ref{tab:boundarycfd}, respectively. Additionally, some examples of 2D meshes are displayed in Figure \ref{fig:meshes}.

\newcolumntype{P}[1]{>{\raggedright\arraybackslash}p{#1}}

\begin{table}[h!]
    \centering
    \captionof{table}{CFD setup per case. Density is $\rho = 1$.}
    \label{tab:setupcfd}
    
    \begin{tabular}{lP{2.2cm}lll}
        \toprule
        \textbf{Case} &
        \textbf{Velocity} &
        \textbf{$\nu$} &
        \textbf{Reynolds} \\
        \midrule
        
        Poiseuille & $u_{max} = 1$ & $10$ & $Re = \dfrac{H}{10}$ \\[7pt]
        
        Channel & $u = 1$ & $2 \times 10^{-2}$ & $Re = 50H$ \\[7pt]
        
        Plate (flat plate) & $u = 1$ & $10^{-2}$ & $Re = 100x$ \\[4pt]
        
        ConvDiff & $u = 1/\sqrt{2}$, $v = u$ & $10$ & $Re = \dfrac{H}{10}$ \\[6pt]
        
        \bottomrule
    \end{tabular}
\end{table}

\begin{table}[h!]
    \centering
    \captionof{table}{Boundary conditions per case (axes as in the dataset).}
    \label{tab:boundarycfd}
    
    \begin{tabular}{l P{2.5cm} P{2.8cm} P{2.8cm} P{2.8cm}}
        \toprule
        \textbf{Boundary} &
        \textbf{Poiseuille} &
        \textbf{Channel} &
        \textbf{Plate } &
        \textbf{ConvDiff} \\
        \midrule
        
        $x^{-}$ &
        Velocity (parabolic) &
        Velocity $(1,0,0)$ &
        Velocity $(1,0,0)$ &
        Inflow $u = 0.705$ \\[18pt]
        
        $x^{+}$ &
        Pressure (fixed) &
        Pressure (outlet) &
        Outflow (zero-gradient) &
        Pressure (open) \\[18pt]
        
        $y^{-}$ &
        Wall (no-slip) &
        Wall (no-slip) &
        Wall (no-slip) &
        Inflow $v = 0.705$ \\[18pt]
        
        $y^{+}$ &
        Wall (no-slip) &
        Wall (no-slip) &
        Pressure (open) &
        Pressure (open) \\[8pt]
        
        \bottomrule
    \end{tabular}
\end{table}

\clearpage

\end{document}